\RequirePackage{silence}
\documentclass[fleqn,usenatbib,useAMS]{mnras} 

\usepackage{newtxtext,newtxmath}

\usepackage{graphicx}
\usepackage{amsmath}

\DeclareMathOperator{\Cov}{Cov}

\usepackage{amsfonts}
\usepackage{float}
\usepackage{bm}
\usepackage{ae,aecompl}
\usepackage{array}
\usepackage{soul}
\usepackage{mathtools}
\usepackage{multirow}
\usepackage[T1]{fontenc}
\usepackage[utf8]{inputenc}
\usepackage{booktabs}
\usepackage{graphicx}
\usepackage{subcaption}
\usepackage[normalem]{ulem}
\usepackage{anyfontsize}
\usepackage{xcolor}

\DeclareRobustCommand{\VAN}[3]{#2}
\let\VANthebibliography\thebibliography
\def\thebibliography{\DeclareRobustCommand{\VAN}[3]{##3}\VANthebibliography}

\DeclareRobustCommand{\appropto}{\mathrel{\vcenter{
		\offinterlineskip\halign{\hfil$##$\cr 
			\propto\cr\noalign{\kern2pt}\sim\cr\noalign{\kern-2pt}}}}}

\makeatletter
\DeclareRobustCommand*\bigcdot{\mathpalette\bigcdot@{2.4}}
\DeclareRobustCommand*\bigcdot@[2]{\mathbin{\vcenter{\hbox{\scalebox{#2}{$\m@th#1\bullet$}}}}}
\makeatother

\makeatletter
\DeclareRobustCommand*\diamond{\mathpalette\diamond@{1.6}}
\DeclareRobustCommand*\diamond@[2]{\mathbin{\vcenter{\hbox{\scalebox{#2}{$\m@th#1\blacklozenge$}}}}}
\makeatother

\defcitealias{Haslbauer_2020}{HBK20}
\defcitealias{Jia_2023}{JHW23}
\defcitealias{Jia_2025}{JHW25}

\title[Testing the local void scenario with BAO data]{Testing the local void hypothesis using baryon acoustic oscillation measurements over the last twenty years} 

\author[I. Banik and V. Kalaitzidis]{Indranil Banik$^{1, 2}$\thanks{E-mail: \href{mailto:indranil.banik@port.ac.uk}{indranil.banik@port.ac.uk} (Indranil Banik); \newline \hspace*{3em} \href{mailto:vk44@st-andrews.ac.uk}{vk44@st-andrews.ac.uk} (Vasileios Kalaitzidis)} and Vasileios Kalaitzidis$^{2}$\vspace{10pt} \\
$^{1}$Institute of Cosmology and Gravitation, University of Portsmouth, Dennis Sciama Building, Burnaby Road, Portsmouth PO1 3FX, UK\\
$^{2}$Scottish Universities Physics Alliance, University of Saint Andrews, North Haugh, Saint Andrews, Fife, KY16 9SS, UK}

\date{Accepted XXX. Received YYY; in original form ZZZ}

\pubyear{\the\year{}}

\begin{document}
\label{firstpage}
\pagerange{\pageref{firstpage}--\pageref{lastpage}}
\maketitle

\begin{abstract} 
A promising solution to the Hubble tension is a local void that is roughly 20\% underdense out to 300~Mpc, as suggested by galaxy number counts in the near-infrared. Gravitationally driven outflows from this KBC void might inflate redshifts enough to solve the Hubble tension, a scenario explored in detail by Haslbauer et al. We obtain predictions for the baryon acoustic oscillation (BAO) observables in their best-fitting void models and in the homogeneous \emph{Planck} cosmology. We compare these models against our compilation of available BAO measurements from the past twenty years. We find that the quality and quantity of available measurements are best using the isotropically averaged distance $D_{\mathrm{V}}$. Taking its ratio with the expected value in the homogeneous model yields good agreement with unity at high redshift, but a discrepancy appears that systematically grows with decreasing redshift. Assuming independent uncertainties, the 42 considered $D_{\mathrm{V}}$ observations give a total $\chi^2$ of 75.7 for the void-free model, while the void models give only $47.3 - 51.2$ depending on the density profile. This represents a reduction in overall tension from $3.3\sigma$ without a void to $1.1\sigma - 1.4\sigma$ in the void models. The $\chi^2$ differences are smaller when considering measurements of the angular BAO scale or its redshift depth. The void-free model provides the worst fit in almost every case. Overall, our results suggest that recent evidence of BAO observables deviating from expectations in the homogeneous \emph{Planck} cosmology could indicate a local void, which was motivated by considerations unrelated to BAO data or the Hubble tension.

\end{abstract}

\begin{keywords}
    cosmological parameters -- cosmology: theory -- cosmology: observations -- distance scale -- large-scale structure of Universe -- gravitation
\end{keywords}

\section{Introduction}
\label{Introduction}

Much of our information about cosmology comes from the cosmic microwave background \citep[CMB;][]{Dicke_1965, Penzias_1965}. Prior to its emission at redshift $z = 1090$, matter and radiation were tightly coupled because of the high Thomson scattering cross-section of free electrons $-$ neutral atoms were not yet stable. Sound waves propagating in the plasma imprinted a characteristic pattern of anisotropies in the CMB, with fluctuations at particular angular scales having more power than at slightly smaller or larger scales \citep{Bernardis_2000, Spergel_2003, Spergel_2007}. This is because perturbations on smaller scales oscillated on a shorter timescale, so perturbations at some particular scales underwent a half-integer number of oscillations in the $\approx 380$~kyr between the Big Bang and recombination. The first acoustic peak in the CMB corresponds to the largest scale with such a local maximum, which in this case is also the global maximum of the whole power spectrum. Its physical size is set by how far a sound wave travelled through the coupled baryon-photon plasma in the limited time before recombination.

The very pronounced first acoustic peak in the CMB implies an excess clustering of matter on a particular scale. This should also be evident at late times in the large-scale matter distribution. A major advance in cosmology has been the reliable detection of precisely this baryon acoustic oscillation (BAO) feature in the large-scale distribution of galaxies \citep{Cole_2005, Eisenstein_2005}. However, the effect is greatly diminished compared to the first peak in the CMB power spectrum. In the standard cosmological model known as Lambda-Cold Dark Matter \citep*[$\Lambda$CDM;][]{Efstathiou_1990, Ostriker_Steinhardt_1995}, this is because baryons fell into the potential wells generated by the dominant CDM component, which lacked oscillatory features in its power spectrum because by definition it did not interact with radiation. Since most of the matter is CDM rather than baryons, the significant oscillations evident in the baryonic power spectrum at recombination were substantially damped at later times.

Although the BAO feature in the galaxy distribution is rather subtle compared to that in the CMB, observations of BAOs at late times do play a very important role in modern cosmology \citep[for a review, see][]{Chen_2024}. This is because the BAO corresponds to a size of about 150 comoving Mpc (cMpc), which is so large that the density fluctuations are rather small, putting them well into the `linear regime' of structure formation. Consequently, although the power on the BAO scale has grown, there has not been much change to the overall shape of the power spectrum plotted with respect to comoving separation. As a result, the length-scale of the BAO peak serves as a standard ruler with fixed comoving length $r_{\mathrm{d}}$ since shortly after recombination. The observed angular scale of the BAO ruler at some redshift tells us the comoving distance to that redshift. We can also gain complementary information from the redshift depth of the BAO ruler, which corresponds to orienting it parallel to the line of sight rather than within the plane of the sky. We discuss the BAO observables in more detail in Section~\ref{BAO_observables_homogeneous}.

Measurements of the cosmic expansion history have become increasingly important in recent years because of the Hubble tension, a persistent discrepancy between estimates of $H_0$ from observations in the early versus late Universe \citep[for a review, see][]{Valentino_2021_problem}. Sometimes referred to as the Hubble constant despite its drastic variation with time, $H_0$ is the present value of the Hubble parameter $H \equiv \dot{a}/a$, where an overdot denotes a derivative with respect to time $t$ since the Big Bang, 0 subscripts indicate present values, and $a$ is the cosmic scale factor normalized to unity today. We can predict $H_0$ by calibrating the $\Lambda$CDM model parameters to the CMB power spectrum, which requires that $H_0 = H_0^{\mathrm{Planck}} = 67.4 \pm 0.5$~km~s$^{-1}$~Mpc$^{-1}$ \citep{Planck_2020, Tristram_2024}. It is also possible to obtain $H_0$ using a variety of other probes that do not rely on modelling the CMB power spectrum, for instance using the ages of the oldest stars and stellar populations in the Galactic disc and halo \citep{Cimatti_2023, Valcin_2025, Xiang_2025}, or using the age difference between stellar populations at different $z$ \citep{Cogato_2024, Guo_2025}. The results are in good agreement with the \emph{Planck} cosmology \citep{Banik_2025}. This is non-trivial given the very narrow region of overlap between their considered non-CMB constraints in the space of $H_0$ and $\Omega_{\mathrm{M}}$, the present fraction of the cosmic critical density accounted for by matter.

We can also obtain $H_0$ from the local redshift gradient $z' \equiv dz/dr$, the rate at which the redshift of extragalactic objects increases with their distance $r$. In a homogeneously expanding universe, the redshift arises entirely from cosmological expansion over the light travel time $r/c$, in which case we must have that $H_0 = cz'$, where $c$ is the speed of light \citep*[see equation~3 of][]{Mazurenko_2025}. The Universe does appear to be homogeneous and isotropic on Gpc scales, where the radio dipole is consistent with the expectation from the Solar velocity \citep[known from the CMB dipole;][]{Planck_2014_Doppler} causing sources ahead of the Sun to appear somewhat brighter, thus increasing the number of detectable sources in that direction \citep[and vice versa; see][]{Wagenveld_2024}. However, $cz'$ is usually measured on scales of only a few hundred Mpc so that $\dot{a}$ changes little over the corresponding range of lookback times \citep[a typical choice is to use $z = 0.023-0.15$ and treat $a \left( t \right)$ as a parabola; see][]{Camarena_2020a, Camarena_2020b}. The measured $cz' = 73.2 \pm 0.9$~km~s$^{-1}$~Mpc$^{-1}$ based on the traditional Cepheid-SNe route \citep{Breuval_2024}, with 4 anchor galaxies for the Cepheid period--luminosity relation and 42 Type~Ia supernovae (SNe) with Cepheid host distances \citep{Riess_2022_comprehensive}. This unexpectedly high local $cz'$ is evident in a wide variety of studies that measure extragalactic distances using different instruments and techniques, some of which rely on SNe while others do not \citep[][and references therein]{Scolnic_2023}. Distances obtained by the \emph{James Webb Space Telescope} are in good agreement with those obtained previously by the \emph{Hubble Space Telescope} \citep{Freedman_2024, Riess_2024_consistency}. Observations of the fundamental plane (FP) relation of elliptical galaxies have recently anchored $cz'$ to $d_{\mathrm{Coma}}$, the distance to the Coma Cluster \citep{Said_2025}. The FP-based $cz'$ is in agreement with other measurements if we adopt the typically reported $d_{\mathrm{Coma}} \approx 100$~Mpc, but $cz' = H_0^{\mathrm{Planck}}$ only if $d_{\mathrm{Coma}} > 110$~Mpc, which exceeds any published measurement since 1990 \citep[][and references therein]{Scolnic_2025}.

A variety of solutions have been proposed for the Hubble tension \citep[][and references therein]{Valentino_2021_solutions, Valentino_2025}. Solutions that modify the physics prior to recombination are disfavoured for several reasons \citep{Vagnozzi_2023}, including difficulties fitting the CMB anisotropies \citep{Calabrese_2025} and the age issue mentioned above \citep*[see also section~7.5 of][]{Poulin_2023}. The impact on the cosmic age can be mitigated if the expansion history is altered only at late times so that $\dot{a}$ is currently about 9\% higher than in the \emph{Planck} cosmology \citep*{Harko_2023, Yao_2023, Rezazadeh_2024}. The required more rapid acceleration of the expansion rate at late times implies an effective phantom equation of state for dark energy \citep{Dahmani_2023}, which may be problematic \citep{Sen_2008, Ludwick_2017}.

These difficulties motivate us to reconsider the assumption that $cz' = \dot{a}$, which is at the heart of the Hubble tension. Standard physics provides two main non-cosmological sources of redshift: peculiar velocity and gravitational redshift (GR). This raises the possibility that the anomalously high local $cz'$ arises from our location within a large and deep void, outflows from which would create extra redshift. Moreover, if we were located close to the void centre, we would be on a potential hill, creating GR contributions to the redshift. Cosmic variance in the local $cz'$ is only 0.9~km~s$^{-1}$~Mpc$^{-1}$ in $\Lambda$CDM \citep{Camarena_2018}, implying that in this paradigm, we cannot be living inside a sufficiently large and deep void to solve the Hubble tension \citep{Wu_2017}. However, it is possible that structure formation is more efficient on scales $\ga 100$~Mpc. This scenario was considered in some detail by \citealt*{Haslbauer_2020} (hereafter \citetalias{Haslbauer_2020}), who considered a simple model of spherically symmetric outflow from a void plus a systemic velocity of the whole void. Those authors evolved voids with three different initial underdensity profiles (Exponential, Gaussian, and Maxwell-Boltzmann) in a background \emph{Planck} cosmology starting from $z = 9$. To enhance the growth of structure, they used a force law based on Milgromian dynamics \citep[MOND;][]{Milgrom_1983, Famaey_McGaugh_2012, Banik_Zhao_2022}.

A local void introduces additional contributions to the redshift. In particular, equation~52 of \citetalias{Haslbauer_2020} shows that the observed redshift $z$ satisfies:
\begin{eqnarray}
    1 + z ~=~ \frac{1}{a \left( t \right)} \overbrace{\sqrt{\frac{c + v_{\mathrm{int}}}{c - v_{\mathrm{int}}}}}^{\text{Doppler}}  \overbrace{\exp \left( \frac{1}{c^2} \int g_{\mathrm{void}} \, dr \right)}^{\text{GR}} \, ,
    \label{z_contributions}
\end{eqnarray}
where $a$ was the cosmic scale factor at the emission time $t$ of an observed photon, $g_{\mathrm{void}}$ is the outward gravitational field induced by the void due to it having less density than the cosmic mean, and $v_{\mathrm{int}}$ is the outflow velocity in the reference frame of the void, which may be moving as a whole. However, the effect of any such systemic velocity would cancel out if observing sources at the same $z$ over a wide range of sky directions, so we do not consider the void's systemic velocity further and deal only with $v_{\mathrm{int}}$. For similar reasons, we also neglect our slight offset from the void centre. These assumptions were also made by \citet{Mazurenko_2025} when assessing how quickly the apparent cosmology returns to the \emph{Planck} cosmology at high redshift, which is inevitable in any local or late-time solution to the Hubble tension. We note that when dealing with observations at very low $z \la 0.05$, we cannot neglect either the systemic velocity of the void or our offset with respect to its centre. Therefore, both of these were carefully considered by \citet{Mazurenko_2024} when comparing the velocity field in the void model against the observed bulk flow curve \citep{Watkins_2023}, which measures the average velocity of galaxies within concentric spheres centred on our location \citep*{Nusser_2016, Peery_2018}.

Galaxy number counts in the near-infrared do indeed reveal a significant underdensity known as the KBC void \citep*{Keenan_2013}. This is an underdense region on a scale of around 300~Mpc \citep[see their figure~11 and figure~1 of][]{Kroupa_2015}. The void is evident at optical \citep{Maddox_1990, Shanks_1990}, infrared \citep{Huang_1997, Busswell_2004, Frith_2003, Frith_2005, Frith_2006, Keenan_2013, Whitbourn_2014, Whitbourn_2016, Wong_2022}, X-ray \citep{Bohringer_2015, Bohringer_2020}, and radio wavelengths \citep*{Rubart_2013, Rubart_2014}. The observed density contrast of the KBC void implies that it would raise the local $cz'$ by about $11 \pm 2\%$, suggesting that the observed local $cz'$ is quite consistent with $\dot{a} = H_0^{\mathrm{Planck}}$ \citepalias[see equation~5 of][]{Haslbauer_2020}. Those authors also found that the void is in $6.04\sigma$ tension with the $\Lambda$CDM model based on a comparison with results from the Millennium XXL simulation \citep[MXXL;][]{Angulo_2012}. The analysis of \citetalias{Haslbauer_2020} indicates that the KBC void is roughly 20\% underdense out to 300~Mpc, but the apparent density contrast is roughly doubled by outflows induced by the void distorting the relation between distance and redshift compared to a homogeneous universe. This causes observers to overestimate their sampled volume when taking data out to some fixed redshift, causing them to find fewer galaxies than they are expecting. The apparent underdensity is $46 \pm 6\%$ without correcting for this effect, as shown by the light blue point on figure~11 of \citet{Keenan_2013}. A somewhat smaller apparent density contrast was claimed by \citet{Wong_2022}, but their figure~1b shows that the comoving luminosity density is still rising by $z = 0.08$, where their measurements stop. This corresponds to about 350~Mpc on figure~11 of \citet{Keenan_2013}, which shows that the luminosity density is nearly flat beyond this point out to the most distant measurements at around 800~Mpc. Therefore, the results in figure~1 of \citet{Wong_2022} should be renormalized to the luminosity density at their last measured point, which is far more representative of the cosmic average. This would cause all their plotted luminosity densities to be scaled down by a factor of $2/3$. With this adjustment, their reported average density within the void of $0.8\times$ the cosmic mean would become only $0.53\times$ the cosmic mean, almost exactly in line with \citet{Keenan_2013}. This highlights the issue that observations with limited depth can underestimate the depth of a local void if they cannot adequately probe beyond it, which is required to correctly normalize the results.

The velocity field predicted by the local supervoid solution to the Hubble tension was further explored by \citet{Mazurenko_2024} using the predicted bulk flow, the average velocity within some radius centred on our location. If we place ourselves at some locations in the proposed void, it is possible to obtain a good match to the observed bulk flow curve \citep{Watkins_2023}, which is based on the CosmicFlows-4 catalogue of galaxy redshifts and redshift-independent distances obtained using various scaling laws \citep{Tully_2023_CF4}. On their largest probed scale of 250~$h^{-1}$~Mpc (where $h \approx 0.7$ is the value of $H_0$ in units of 100~km~s$^{-1}$~Mpc$^{-1}$), the observed bulk flows are quadruple the $\Lambda$CDM expectation and thus in $>5\sigma$ tension with it \citep[see figure~8 of][]{Watkins_2023}. It was later found that their reported bulk flows are in ``excellent agreement'' with those found by \citet*{Whitford_2023} out to 173~$h^{-1}$~Mpc, the outer limit to their more conservative analysis.

In this contribution, we test the local void model further out using a compilation of BAO measurements collected over the last twenty years. Outflow from a local void and GR due to our location on a potential hill inflate the redshift to any fixed distance, slightly distorting the relation between redshift and the BAO observables (Section~\ref{BAO_observables_void}). We predict these observables for three different initial void underdensity profiles (Exponential, Gaussian, and Maxwell-Boltzmann), using the best-fitting parameters for each profile as published in tables~4 and C1 of \citetalias{Haslbauer_2020}. Those authors mainly constrained the void parameters using the local $cz'$ from SNe and the density profile of the KBC void, while fixing the background expansion history to the \emph{Planck} cosmology to ensure consistency with the CMB (Appendix~\ref{CMB_impact_void}). Importantly, they did not consider BAO data, much of which was in any case not available at that time. This allows BAO observations to provide an independent test of their model.

As part of this test, we can also address whether a local void might be responsible for hints that recent BAO observations with the Dark Energy Spectroscopic Instrument \citep[DESI;][]{DESI_2016, DESI_2022} deviate from expectations in the homogeneous \emph{Planck} cosmology \citep{DESI_2024}. These hints for an anomaly primarily come from two redshift bins, so one has to exclude a significant amount of data to remove the hints \citep{Wang_2024_BAO}. Moreover, there were indications from BAO measurements prior to DESI that $H \left( z \right)$ rises above expectations in the \emph{Planck} cosmology at low $z$ \citep{Gomez_2024}. The DESI anomaly has been interpreted as indicating time variation of the dark energy density \citep{Giare_2024, Wang_2024_BAO}, but we explore if it might instead be due to a local void.

After explaining our methods in Section~\ref{Methods}, we present our results in Section~\ref{Results} and discuss them in Section~\ref{Discussion}. We then conclude in Section~\ref{Conclusions}. We use $D_{\mathrm{c}}$ to denote the comoving distance, which some studies call $D_{\mathrm{M}}$.

\section{Methods}
\label{Methods}

We explain how the BAO observables should be determined in a homogeneous cosmology (Section~\ref{BAO_observables_homogeneous}), for which we adopt the parameters $H_0 = 67.4$~km~s$^{-1}$~Mpc$^{-1}$ and $\Omega_{\mathrm{M}} = 0.315$ \citep{Planck_2020}. These values were adopted by \citetalias{Haslbauer_2020}, with a recent analysis of the final \emph{Planck} data release barely affecting the results \citep{Tristram_2024}. We then discuss how the above predictions for the BAO observables should be revised to account for a local void (Section~\ref{BAO_observables_void}). We compare the homogeneous and local void models using a compilation of BAO observations over the last twenty years, which we discuss in Section~\ref{BAO_sample_section}.

\subsection{The BAO observables in a homogeneous cosmology}
\label{BAO_observables_homogeneous}

Since the BAO feature is a 3D structure, we can detect it in the correlation function of galaxies both within the plane of the sky and along the line of sight. The transverse BAO signal appears as excess power at a particular angular scale
\begin{eqnarray}
    \Delta \theta ~=~ \frac{r_{\mathrm{d}}}{D_{\mathrm{c}}} \, ,
    \label{Delta_theta_BAO}
\end{eqnarray}
where $D_{\mathrm{c}}$ is the comoving distance to the effective redshift of the galaxies used to measure the BAO signal. $D_{\mathrm{c}}$ can be found from the underlying expansion history $a \left( t \right)$ using
\begin{eqnarray}
    D_{\mathrm{c}} ~=~ \int_t^{t_0} \frac{c \, dt'}{a \left( t' \right)} \, = \, \int_0^{z} \frac{c \, dz'}{H \left( z' \right)} \, ,
    \label{D_c}
\end{eqnarray}
where $t$ is the time since the Big Bang when the galaxies are observed, $t_0$ is the present epoch, and $z = a^{-1} - 1$ is the redshift. Note that if we were to calculate $\Delta \theta$ using instead the physical ruler size $a r_{\mathrm{d}}$, we would get this in the numerator of Equation~\ref{Delta_theta_BAO}, but we would then need to use the angular diameter distance $D_{\mathrm{A}} = a D_{\mathrm{c}}$ in the denominator, causing the factors of $a$ to cancel.

The parallel BAO signal appears as excess power at a particular difference between the redshifts of galaxies within the considered shell. The redshift depth of the BAO ruler is
\begin{eqnarray}
    \Delta z ~=~ \frac{r_{\mathrm{d}}}{D_{\mathrm{H}}} = \frac{r_{\mathrm{d}} H\left( z \right)}{c} \, ,
    \label{Delta_z_BAO}
\end{eqnarray}
with the comoving Hubble distance $D_{\mathrm{H}} \equiv c/H$ at any $z$. It will be helpful to think of $\Delta z$ as arising from the different amounts of cosmic expansion over the different light travel times from the near and far ends of the BAO ruler, orienting it parallel to the line of sight.

\subsection{Predicting the BAO observables from within a local void}
\label{BAO_observables_void}

A local void introduces additional non-cosmological contributions to the redshift (Equation~\ref{z_contributions}). As a result, the integral with respect to $z$ in Equation~\ref{D_c} must be understood as an integral with respect to the purely cosmological contribution to the redshift, which is
\begin{eqnarray}
    z_c ~\equiv~ a^{-1} - 1 \, .
    \label{zc}
\end{eqnarray}
A local void does not affect Equation~\ref{Delta_theta_BAO}, but the additional non-cosmological redshift distorts the relation between $D_{\mathrm{c}}$ and $z$ by raising the latter. This changes the predicted $\Delta \theta \left( z \right)$.

We obtain the predicted BAO observables in the local void models considered by \citetalias{Haslbauer_2020} using the techniques discussed below. Those authors evolved a small initial underdensity forwards from $z = 9$ until the present time assuming three different initial underdensity profiles: Exponential, Gaussian, and Maxwell-Boltzmann. In each case, the authors found best-fitting void parameters (size, depth, and external field strength) by comparing to a variety of observations, including especially the observed density profile of the KBC void and the local $cz'$. However, they did not consider BAO data. In this contribution, we adopt their best-fitting parameters for each void profile as published in their tables~4 and C1, thereby obtaining \emph{a priori} predictions for the BAO observables. We note that since \citetalias{Haslbauer_2020} assumed a background \emph{Plank} cosmology and the underlying idea is that structure growth is enhanced on scales $\ga 100$~Mpc, the CMB anisotropies can be fit just as well as in $\Lambda$CDM (we discuss this further in Appendix~\ref{CMB_impact_void}). We therefore focus on the late Universe.

Since the available BAO measurements (Section~\ref{BAO_sample_section}) are generally quite far out, we approximate that we are located at the void centre. This is justified by the finding that we should be $\la 150$~Mpc from the void centre to match the observed bulk flow curve, with the best match requiring an even smaller distance \citep{Mazurenko_2024}. This allows us to simplify our analysis substantially. However, the large redshifts of BAO measurements also force us to relax their assumption that light travel times are negligible, requiring us to consider our past lightcone in greater detail (Equation~\ref{Lightcone}).

\subsubsection{Angular scale}
\label{Angular_scale}

We begin by finding the predicted $D_{\mathrm{c}} \left( a \right)$ in the homogeneous \emph{Planck} cosmology with parameters $H_0 = 67.4$~km~s$^{-1}$~Mpc$^{-1}$ and $\Omega_{\mathrm{M}} = 0.315$ \citep{Planck_2020} as these were the parameters adopted by \citetalias{Haslbauer_2020}. We note that these parameters have been updated only slightly in the final \emph{Planck} data release \citep{Tristram_2024}.

Similarly to \citetalias{Haslbauer_2020}, we consider a grid of particle trajectories with uniform spacing in their initial separation from the void centre. We find the intersection of each trajectory with our past lightcone using the same approach as in their section~3.3.3, which requires us to solve
\begin{eqnarray}
    r_c \left( t \right) ~=~ \int_t^{t_0} \frac{c \, dt'}{a \left( t' \right)} \, ,
    \label{Lightcone}
\end{eqnarray}
where $r_c$ is the comoving distance of the particle from the void centre, $t$ is the time at which we observe the particle, and $t_0$ is the present time. We solve this for $t$ using the Newton-Raphson method (using also information on the particle velocity) and record both $a$ and $z$ at that time, which we find using equations~45 and 52 of \citetalias{Haslbauer_2020}, respectively. We then find $D_{\mathrm{c}}$ by interpolating to the desired $a$ using our previously found $D_{\mathrm{c}} \left( a \right)$ assuming homogeneity (Equation~\ref{D_c}), which tells us the comoving distance to the particle at time $t$. By repeating this procedure for all considered test particles, we obtain the predicted $D_{\mathrm{c}} \left( z \right)$. We also find the corresponding prediction without any local void by writing $D_{\mathrm{c}} \left(a \right)$ as $D_{\mathrm{c}} \left( z_c \right)$ using Equation~\ref{zc} and assuming that $z = z_c$.

Since the BAO angular scale (Equation~\ref{Delta_theta_BAO}) varies hugely with $z$, it is common to divide $\Delta \theta$ by its value in some fiducial cosmology to obtain the parameter
\begin{eqnarray}
    \alpha_\perp ~\equiv~ \frac{D_{\mathrm{c}}}{r_{\mathrm{d}}} \div \left. \frac{D_{\mathrm{c}}}{r_{\mathrm{d}}} \right|_{\mathrm{fid}} \, ,
    \label{alpha_perp}
\end{eqnarray}
where `fid' subscripts indicate values calculated at the same $z$ in the fiducial cosmology. In our case, this is the homogeneous \emph{Planck} cosmology, which gives $r_{\mathrm{d}} = 147.05 \pm 0.30$~cMpc \citep[section~5.4 of][]{Planck_2020}. Since the uncertainty in $r_{\mathrm{d}}$ is very small, we often discuss our results as if $\alpha_\perp$ measures $D_{\mathrm{c}}$ divided by its fiducial value at the same $z$.

\subsubsection{Redshift depth}
\label{Redshift_depth}

The impact of a local void on $\Delta z$ must be found by considering the redshifts of photons emitted from the near and far ends of the BAO ruler. The difference $\Delta z$ arises from three main effects:
\begin{enumerate}
    \item Cosmological expansion over the light travel time across the BAO ruler;
    \item Peculiar velocity variations within the void; and
    \item Evolution of the void over the light travel time.
\end{enumerate}
To obtain the redshift depth of the BAO ruler at redshift $z$, we start from the comoving distance $D_{c,0}$ predicted by the void model to that $z$, including the non-cosmological contributions to $z$ (Equation~\ref{z_contributions}). We then quantify the redshifts $z_\pm$ at which $D_{\mathrm{c}} = D_{c,0} \pm r_{\mathrm{d}}$. We find $z_\pm$ using the Newton-Raphson algorithm, applying the secant method as it is difficult to analytically quantify the gradient. We then set
\begin{eqnarray}
    \Delta z ~=~ \frac{z_+ - z_-}{2} \, ,
\end{eqnarray}
which we use to calculate $D_{\mathrm{H}}/r_{\mathrm{d}}$ via Equation~\ref{Delta_z_BAO}. We use this centred differencing approach to make our results more accurate. We repeat our calculations including only the cosmological contribution to $z$ (Equation~\ref{zc}) to get the results without a local void.

Similarly to Equation~\ref{alpha_perp}, the redshift depth of the BAO ruler is usually normalized to its value in some fiducial cosmology by defining
\begin{eqnarray}
    \alpha_\parallel ~\equiv~ \frac{D_{\mathrm{H}}}{r_{\mathrm{d}}} \div \left. \frac{D_{\mathrm{H}}}{r_{\mathrm{d}}} \right|_{\mathrm{fid}} \, .
    \label{alpha_parallel}
\end{eqnarray}
Since the uncertainty in the \emph{Planck} $r_{\mathrm{d}}$ is very small, we often discuss our results as if $\alpha_\parallel$ measures $D_{\mathrm{H}}$ divided by its fiducial value at the same $z$.

\subsubsection{The Alcock-Paczynski (AP) test}
\label{AP_test}

Although there is very little uncertainty in the \emph{Planck} value of $r_{\mathrm{d}}$, this is reliant on theoretical assumptions concerning the early Universe. In principle, BAO measurements can provide useful cosmological constraints without assuming any particular value for $r_{\mathrm{d}}$ as long as we make the much less strict assumption that $r_{\mathrm{d}}$ has remained constant with time, a technique known as uncalibrated cosmic standards \citep*[UCS;][]{Lin_2021}.

Altering $r_{\mathrm{d}}$ would scale both $\alpha_\parallel$ and $\alpha_\perp$ by the same factor. This motivates us to consider their ratio, which is known as the AP parameter $\alpha_{_{\mathrm{AP}}}$ \citep{Alcock_1979}:
\begin{eqnarray}
    \alpha_{_{\mathrm{AP}}} ~\equiv~ \frac{\alpha_\perp}{\alpha_\parallel} \, .
    \label{alpha_AP}
\end{eqnarray}
This cancels out the effect of possible issues with the fiducial $r_{\mathrm{d}}$, so that $\alpha_{_{\mathrm{AP}}} \neq 1$ (a failure of the AP test) cannot be explained merely by altering $r_{\mathrm{d}}$ if we assume homogeneity. Instead, we would have to modify the \emph{shape} of the expansion history. For this reason, the AP test was originally proposed as a way to detect dark energy.

A major drawback of $\alpha_{_{\mathrm{AP}}}$ is that obtaining it observationally requires the ratio between two quantities each with their own uncertainty, making it less accurate \citep[though see the top right panel of figure~6 in][]{DESI_2025}.

\subsubsection{Isotropic average}
\label{Isotropic_average}

Though we have only discussed the BAO feature within the plane of the sky and along the line of sight, it is actually a 3D feature. The volume of this feature can be computed using the BAO observables, with the cube root of this volume then providing an isotropically averaged estimate of the size of the BAO ruler. This consideration has motivated the introduction of the derived BAO observable $D_{\mathrm{V}}$ \citep{Eisenstein_2005}, which is defined as follows \citep[see equation~2.6 of][]{DESI_2024}:
\begin{eqnarray}
    D_{\mathrm{V}} ~\equiv~ \left( z D_{\mathrm{c}}^2 D_{\mathrm{H}} \right)^{1/3} \, .
    \label{D_V}
\end{eqnarray}

Similarly to the definitions of $\alpha_\perp$ and $\alpha_\parallel$, it is common to divide $D_{\mathrm{V}}$ by its predicted value in some fiducial cosmology at the same $z$, which cancels out the factor of $z$ in Equation~\ref{D_V}. The ratio of $D_{\mathrm{V}}$ to its fiducial value is thus
\begin{eqnarray}
    \alpha_{\mathrm{iso}} ~\equiv~ \frac{D_{\mathrm{V}}}{r_{\mathrm{d}}} \div \left. \frac{D_{\mathrm{V}}}{r_{\mathrm{d}}} \right|_{\mathrm{fid}} ~=~ \left( \alpha_\perp^2 \alpha_\parallel \right)^{1/3} \, ,
    \label{alpha_iso}
\end{eqnarray}
where the `iso' subscript indicates an isotropically averaged value.

A major advantage of $\alpha_{\mathrm{iso}}$ is that since it is basically the geometric mean of $\alpha_\perp$ and $\alpha_\parallel$ (with the former counted twice), uncertainties are reduced for much the same reason that the average of two measurements is more accurate than either measurement on its own. This allows observers to report $D_{\mathrm{V}}$ even if the results are too inaccurate to report $D_{\mathrm{c}}$ and $D_{\mathrm{H}}$ individually.

\subsection{The BAO sample}
\label{BAO_sample_section}

We test the local void scenario using a compilation of published BAO measurements from the last twenty years. These studies are summarized in Table~\ref{BAO_sample_table}, sorted in ascending order of redshift. Many studies are listed multiple times because they contribute results at different redshifts. At each redshift, we indicate which distance measures were published in that study.

\begin{table*}
    \centering
    \caption{Our compilation of available BAO measurements from the indicated survey and reference, sorted in ascending order of redshift. Each symbol in columns $4-7$ indicates that the cited study was used to obtain the BAO observable given in the column heading. Circle symbols indicate that we used the published value, while diamond symbols indicate that we calculated the BAO observable using the information provided. Black symbols indicate values used in our $\chi^2$ calculation (Section~\ref{chi_sq_analysis}), while grey symbols indicate values not used there. This occurs if the results were later reanalysed and used for a different data point, or if results are available using both coarser and finer redshift bins, in which case we plot the former to avoid crowding but use the latter for $\chi^2$ to extend the redshift range. We provide brief comments where necessary in the final column.}
    \begin{tabular}{cccccccc}
        \hline
        $z$ & Survey & Reference & $D_{\mathrm{c}}$ & $D_{\mathrm{H}}$ & $\alpha_{_{\mathrm{AP}}}$ & $D_{\mathrm{V}}$ & Notes \\ [2pt] \hline
        0.068 & Ho'oleilana & \citet{Tully_2023_BAO} & & & & $\textcolor{gray}{\bigcdot}$ & \raisebox{-0.3ex}{$\alpha_{\mathrm{iso}}$ from individual structure at $z = 0.068^{+0.003}_{-0.007}$} \\ [2pt]
        0.097 & 6dFGS & \citet{Carter_2018} & & & & $\textcolor{gray}{\bigcdot}$ & $r_{\mathrm{d}} = 147.5$ cMpc  \\
        0.106 & 6dFGS & \citet{Beutler_2011} & & & & $\bigcdot$ & \\
        0.11 & SDSS DR17 & \citet{Carvalho_2021} & $\bigcdot$ & & & & \\
        0.122 & SDSS + 6dFGS & \citet{Carter_2018} & & & & $\bigcdot$ & $r_{\mathrm{d}} = 147.5$ cMpc \\
        0.15 & SDSS DR7 & \citet{Ross_2015} & & & & \textcolor{gray}{$\bigcdot$} & $r_{\mathrm{d}} = 148.69$ cMpc \\
        0.2 & SDSS + 2dFGRS & \citet{Percival_2007} & & & & $\bigcdot$ & \\
        0.2 & SDSS DR7 & \citet{Percival_2010} & & & & $\bigcdot$ & \\
        0.24 & SDSS DR12 & \citet{Chuang_2017} & $\bigcdot$ & $\bigcdot$ & $\diamond$ & $\bigcdot$ & $r_{\mathrm{d}} = 147.66$ cMpc; finer redshift bin \\
        0.295 & DESI DR2 & \citet{DESI_2025} & & & & $\bigcdot$ & \\
        0.32 & SDSS DR9 & \citet{Anderson_2014} & & & & $\bigcdot$ & $r_{\mathrm{d}} = 149.28$ cMpc \\
        0.32 & SDSS DR12 & \citet{Alam_2017} & $\bigcdot$ & $\bigcdot$ & $\diamond$ & $\diamond$ & \\
        0.32 & SDSS DR12 & \citet{Chuang_2017} & $\textcolor{gray}{\bigcdot}$ & $\textcolor{gray}{\bigcdot}$ & $\textcolor{gray}{\diamond}$ & $\textcolor{gray}{\bigcdot}$ & Coarser redshift bin \\
        0.35 & SDSS DR5 & \citet{Percival_2007} & & & & $\bigcdot$ & \\
        0.35 & SDSS DR7 & \citet{Percival_2010} & & & & $\bigcdot$ & \\
        0.35 & SDSS DR7 & \citet*{Chuang_2012a} & & & & $\bigcdot$ & \\
        0.35 & SDSS DR7 & \citet{Chuang_2012b} & & & & $\bigcdot$ & \\
        0.37 & SDSS DR12 & \citet{Chuang_2017} & $\bigcdot$ & $\bigcdot$ & $\diamond$ & $\bigcdot$ & Finer redshift bin \\
        0.38 & SDSS DR12 & \citet{Alam_2017} & $\bigcdot$ & $\bigcdot$ & $\diamond$ & $\diamond$ & $r_{\mathrm{d}} = 147.78$ cMpc \\
        0.38 & SDSS DR12 & \citet{Ivanov_2020} & $\bigcdot$ & $\bigcdot$ & $\diamond$ & $\diamond$ & $r_{\mathrm{d}} = 147.09$ cMpc \\
        0.38 & SDSS DR12 & \citet*{Schirra_2024} & $\bigcdot$ & $\bigcdot$ & $\diamond$ & $\bigcdot$ & \\
        0.44 & WiggleZ Final DR & \citet{Blake_2012} & $\bigcdot$ & $\bigcdot$ & $\diamond$ & $\bigcdot$ & \\
        0.49 & SDSS DR12 & \citet{Chuang_2017} & $\bigcdot$ & $\bigcdot$ & $\diamond$ & $\bigcdot$ & Finer redshift bin \\
        0.51 & SDSS DR12 & \citet{Alam_2017} & $\bigcdot$ & $\bigcdot$ & $\diamond$ & $\diamond$ & \\
        0.51 & DESI DR2 & \citet{DESI_2025} & $\bigcdot$ & $\bigcdot$ & $\bigcdot$ & $\bigcdot$ & \\
        0.54 & SDSS DR8 & \citet{Seo_2012} & $\bigcdot$ & & & & \\
        0.57 & SDSS DR12 & \citet{Alam_2017} & $\bigcdot$ & $\bigcdot$ & $\diamond$ & $\diamond$ & $r_{\mathrm{d}} = 147.78$ cMpc \\
        0.57 & SDSS DR12 & \citet{Slepian_2017} & & & & $\bigcdot$ & $r_{\mathrm{d}} = 147.66$ cMpc \\
        0.59 & SDSS DR12 & \citet{Chuang_2017} & $\textcolor{gray}{\bigcdot}$ & $\textcolor{gray}{\bigcdot}$ & $\textcolor{gray}{\diamond}$ & $\textcolor{gray}{\diamond}$ &  Coarser redshift bin \\
        0.6 & WiggleZ Final DR & \citet{Blake_2012} & $\bigcdot$ & $\bigcdot$ & $\diamond$ & $\bigcdot$ & $r_{\mathrm{d}} = 153.3$ cMpc \citep{Blake_2011} \\
        0.61 & SDSS DR12 & \citet{Alam_2017} & $\bigcdot$ & $\bigcdot$ & $\diamond$ & $\bigcdot$ & $r_{\mathrm{d}} = 147.78$ cMpc \\
        0.61 & SDSS DR12 & \citet{Ivanov_2020} & $\bigcdot$ & $\bigcdot$ & $\diamond$ & $\diamond$ & $r_{\mathrm{d}} = 147.09$ cMpc \\
        0.61 & SDSS DR12 & \citet{Schirra_2024} & $\bigcdot$ & $\bigcdot$ & $\diamond$ & $\bigcdot$ & \\
        0.64 & SDSS DR12 & \citet{Chuang_2017} & $\bigcdot$ & $\bigcdot$ & $\diamond$ & $\bigcdot$ & Finer redshift bin \\
        0.7 & DECaLS DR8 & \citet{Sridhar_2020} & $\bigcdot$ & & & & $r_{\mathrm{d}} = 147.05$ cMpc \citep{Planck_2020} \\
        0.7 & SDSS DR16 & \citet{Hector_2020} & $\bigcdot$ & $\bigcdot$ & $\diamond$ & $\diamond$ & \\
        0.7 & SDSS DR16 & \citet{Zhao_2021} & $\bigcdot$ & $\bigcdot$ & $\diamond$ & $\diamond$ & Results identical to 2020 preprint \\
        0.706 & DESI DR2 & \citet{DESI_2025} & $\bigcdot$ & $\bigcdot$ & $\bigcdot$ & $\bigcdot$ & \\
        0.73 & WiggleZ Final DR & \citet{Blake_2011} & & & & $\bigcdot$ & \\
        0.73 & WiggleZ Final DR & \citet{Blake_2012} & $\bigcdot$ & $\bigcdot$ & $\diamond$ & $\diamond$ & $r_{\mathrm{d}} = 153.3$ cMpc \citep{Blake_2011} \\
        0.77 & SDSS DR16 & \citet{Wang_2020_SDSS} & $\bigcdot$ & $\bigcdot$ & $\diamond$ & $\diamond$ & \\
        0.8 & SDSS DR14 & \citet{Zhu_2018} & $\bigcdot$ & $\bigcdot$ & $\diamond$ & $\diamond$ & $r_{\mathrm{d}} = 147.18$ cMpc \citep{Planck_2020} \\
        0.81 & DES Y1 & \citet{DES_2019} & $\bigcdot$ & & & & \\
        0.835 & DES Y3 & \citet{DES_2022} & $\bigcdot$ & & & & \\
        0.85 & SDSS DR16 & \citet{Mattia_2021} & & & & $\bigcdot$ & \\
        0.85 & DES Y6 & \citet{DES_2024} & $\bigcdot$ & & & & \\
        0.874 & DECaLS DR8 & \citet{Sridhar_2020} & $\bigcdot$ & & & & $r_{\mathrm{d}} = 147.05$ cMpc \citep{Planck_2020} \\
        0.934 & DESI DR2 & \citet{DESI_2025} & $\bigcdot$ & $\bigcdot$ & $\bigcdot$ & $\bigcdot$ & \\
        1.321 & DESI DR2 & \citet{DESI_2025} & $\bigcdot$ & $\bigcdot$ & $\bigcdot$ & $\bigcdot$ & \\
        1.48 & SDSS DR16 & \citet{Neveux_2020} & $\bigcdot$ & $\bigcdot$ & $\diamond$ & $\diamond$ & Consensus BAO results \citep[see also][]{Hou_2021} \\
        1.484 & DESI DR2 & \citet{DESI_2025} & $\bigcdot$ & $\bigcdot$ & $\bigcdot$ & $\bigcdot$ & \\
        2.33 & DESI DR2 & \citet{DESI_2025} & $\bigcdot$ & $\bigcdot$ & $\bigcdot$ & $\bigcdot$ & \\
        2.33 & DESI Y1 + SDSS & \citet{DESI_2024} & $\bigcdot$ & $\bigcdot$ & $\diamond$ & $\diamond$ & DESI Y1 results identical to 2024 preprint \\
        2.33 & SDSS DR16 & \citet{Bourboux_2020} & $\textcolor{gray}{\bigcdot}$ & $\textcolor{gray}{\bigcdot}$ & $\textcolor{gray}{\diamond}$ & $\textcolor{gray}{\diamond}$ & \\
        2.4 & DES Y1 & \citet{DES_SPT_2018} & $\bigcdot$ & $\bigcdot$ & $\diamond$ & $\diamond$ & \\ \hline
    \end{tabular}
    \label{BAO_sample_table}
\end{table*}

In cases where $D_{\mathrm{c}}$ and $D_{\mathrm{H}}$ are both available, we can calculate $D_{\mathrm{V}}$ using Equation~\ref{D_V}. We sometimes do so, but if possible, we use instead the published value of $D_{\mathrm{V}}$ and its uncertainty. If we have to work out the uncertainty ourselves, we assume independent errors on $D_{\mathrm{c}}/r_{\mathrm{d}}$ and $D_{\mathrm{H}}/r_{\mathrm{d}}$ given that these refer to orthogonal directions. We then combine their fractional errors in quadrature to obtain the fractional error on $D_{\mathrm{V}}/r_{\mathrm{d}}$.
\begin{eqnarray}
    \widetilde{\sigma} \left( \frac{D_{\mathrm{V}}}{r_{\mathrm{d}}} \right) ~=~ \sqrt{\left[ \frac{2}{3} \widetilde{\sigma} \left( \frac{D_{\mathrm{c}}}{r_{\mathrm{d}}} \right) \right]^2 + \left[ \frac{1}{3} \widetilde{\sigma} \left( \frac{D_{\mathrm{H}}}{r_{\mathrm{d}}} \right) \right]^2} \, ,
    \label{sigma_D_V}
\end{eqnarray}
where $\widetilde{\sigma} \left( x \right) \equiv \sigma \left( x \right)/x$ is the fractional uncertainty on any quantity $x$ with uncertainty $\sigma \left( x \right)$. We indicate which measurements of $D_{\mathrm{V}}$ were obtained in this way and which can be read off from the cited publication.

We work out the AP parameter in cases where we have both $D_{\mathrm{c}}$ and $D_{\mathrm{H}}$. To get the uncertainty in $\alpha_{_{\mathrm{AP}}}$, we again assume that $D_{\mathrm{c}}$ and $D_{\mathrm{H}}$ have independent fractional uncertainties, which we simply add in quadrature to obtain the fractional uncertainty in $\alpha_{_{\mathrm{AP}}}$. This is similar to Equation~\ref{sigma_D_V}, but with the factors of 1/3 and 2/3 both replaced by 1.

Although the BAO observable is a dimensionless ratio of some distance measure with $r_{\mathrm{d}}$, some studies give the distance measure in Mpc, assuming a particular value for $r_{\mathrm{d}}$. More modern studies are generally careful to avoid making this model-dependent assumption when publishing observations, but some older studies suffer from this issue. Fortunately, such studies generally state their assumed $r_{\mathrm{d}}$, which we indicate in Table~\ref{BAO_sample_table}. In some cases, the assumed value of $r_{\mathrm{d}}$ is not mentioned in the study, but $r_{\mathrm{d}}$ can be obtained through the studies it cites. In these cases, we indicate the assumed $r_{\mathrm{d}}$ and also add a citation, since this does not follow directly from the study giving the BAO measurement. In a very small number of cases, several studies are cited in relation to the assumed $r_{\mathrm{d}}$, making it unclear exactly what value was assumed when reporting the BAO distance measure in Mpc. In these cases, the cited studies all give numerically very similar values of $r_{\mathrm{d}}$, so we pick whichever one we think is most suitable. We caution that older studies that list a BAO distance measure in Mpc often assume a slightly different $r_{\mathrm{d}}$ to the modern estimate of 147~cMpc because they pre-date the \emph{Planck} mission, forcing them to make do with older CMB observations, generally from the \emph{Wilkinson Microwave Anisotropy Probe} \citep[\emph{WMAP};][]{Bennett_2003, Hinshaw_2003, Spergel_2003}. Comparison of these older BAO measurements with a CMB-derived cosmology should be redone in the \emph{Planck} era with updated $r_{\mathrm{d}}$. This can be calculated using equation~6 of \citet{Eisenstein_1998}, with equation~3.4 of \citet*{Brieden_2023} providing an expansion about the \emph{Planck} parameters \citep{Planck_2020}.

\section{Results}
\label{Results}

\begin{figure*}
    \includegraphics[width=\textwidth]{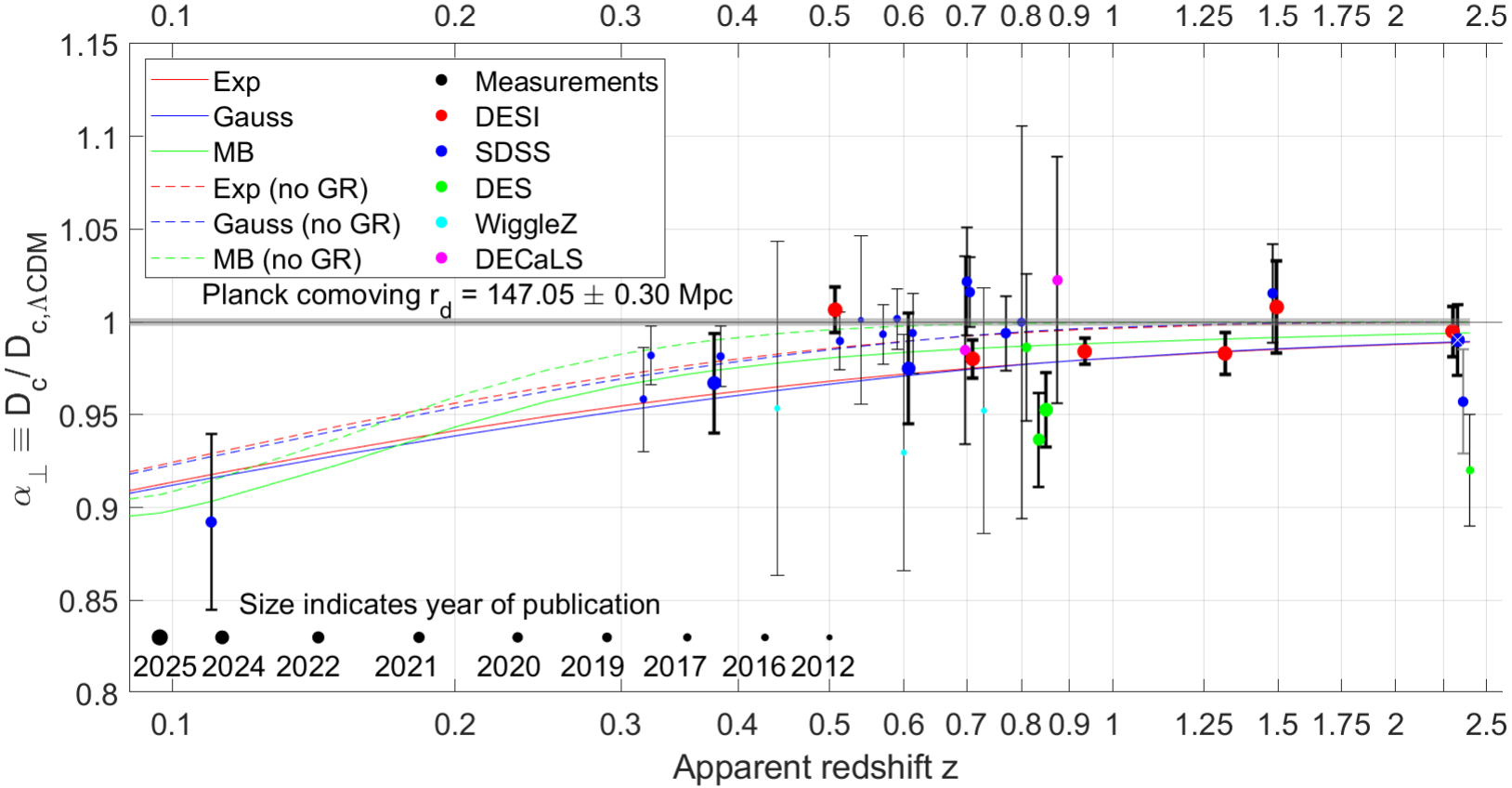}
    \caption{The ratio $\alpha_\perp$ (Equation~\ref{alpha_perp}) of the comoving distance $D_{\mathrm{c}}$ to the $\Lambda$CDM expectation at the same redshift as a function of the apparent redshift $z$, both according to the model (solid lines) and observations (points with uncertainties). Different void profiles are shown in different colours, as indicated in the legend. For illustrative purposes, the dashed lines show the same models without the GR contribution (Equation~\ref{z_contributions}). The observations used here are indicated using a symbol in the $D_{\mathrm{c}}$ column of Table~\ref{BAO_sample_table}. The size of each point and the thickness of its associated error bar have been scaled based on the year of publication, as indicated towards the lower left. The redshifts shown here have been dithered slightly at $z = 0.32$, 0.51, 1.48, and 2.33 to make the points more clearly visible. The point at $\left(z = 2.33, \alpha_\perp = 0.9902 \right)$ with an overlaid white cross is a reanalysis of both the more recent and older observations at a similar $z$, so we show the latter with grey error bars.}
    \label{D_c_graph}
\end{figure*}

We begin by presenting results on $D_{\mathrm{c}}$, which is inversely related to the BAO angular scale $\Delta \theta$ (Equation~\ref{Delta_theta_BAO}). Our results are shown in Figure~\ref{D_c_graph} using the parameter $\alpha_\perp$ (Equation~\ref{alpha_perp}). By definition, the expectation in the homogeneous \emph{Planck} cosmology is 1 at all $z$, though there is a small 0.2\% uncertainty equal to the fractional uncertainty on the \emph{Planck} $r_{\mathrm{d}}$, which we show as a grey band. The coloured solid lines show results from the three void profiles considered by \citetalias{Haslbauer_2020}, adopting their best-fitting void parameters in each case. These parameters were obtained without regard to BAO results, many of which were not available at that time. All three models have $\alpha_\perp \to 1$ at high $z$ because the effect of a local void decays beyond its `edge' \citep[see also][]{Mazurenko_2025}. At low $z$, these models reach $\alpha_\perp \approx 0.9$ because the models were constrained to solve the Hubble tension, implying that $cz' \approx 1.1 \, H_0^{\mathrm{Planck}}$. This means that in the local Universe, the distance to any fixed redshift is about 10\% smaller than in the homogeneous \emph{Planck} cosmology. Since the void models were constrained using the observed density profile of the KBC void \citep{Keenan_2013}, they all converge to unity at high $z$ at a similar rate. This makes it difficult to distinguish the models from each other. We also show results without the GR contribution using dashed lines, helping to show its impact by comparison with solid lines in the same colour. At large redshift, there is negligible outflow velocity due to a local void, making the GR contribution mainly responsible for the predicted deviations from $\alpha_\perp = 1$. This behaviour is also evident in figure~3 of \citet{Mazurenko_2025}.

Our results in Figure~\ref{D_c_graph} show that all of the void models provide a better match to the observations than the homogeneous \emph{Planck} cosmology -- we will quantify this later (Section~\ref{chi_sq_analysis}). The vast majority of the observations lie at $\alpha_\perp < 1$, which is not expected if the latent value is 1 and deviations arise only from random observational errors. More recent and older observations are in good agreement, as can be seen by comparing symbols with different sizes. Results from different surveys are also in good agreement, as can be seen by comparing symbols with different colours. In particular, more recent results from the Sloan Digital Sky Survey \citep[SDSS;][]{SDSS} also generally prefer $\alpha_\perp < 1$. This is in line with similar hints from the DESI results \citep{Wang_2024_BAO}. It is difficult to obtain BAO results at $z \la 0.3$ where a local void would have the greatest impact, due to the large angular size of the BAO feature. The single available $D_{\mathrm{c}}$ measurement in this crucial regime is much more in line with the local void models than with the void-free model.

\begin{figure*}
    \includegraphics[width=\textwidth]{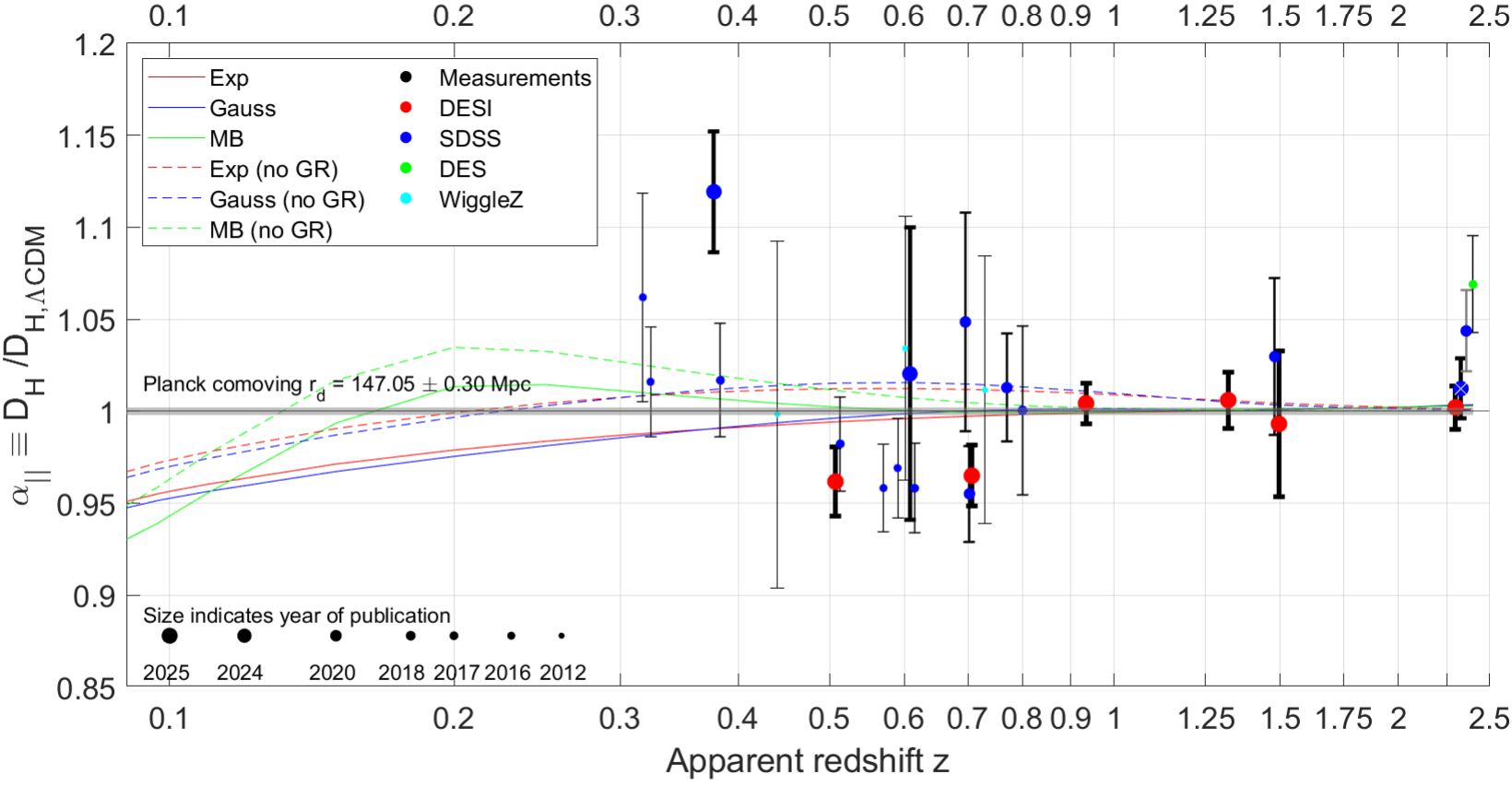}
    \caption{The ratio $\alpha_\parallel$ (Equation~\ref{alpha_parallel}) of the Hubble distance $D_{\mathrm{H}}$ to the $\Lambda$CDM expectation at the same redshift as a function of the apparent redshift $z$, both according to the model (solid lines) and observations (points with uncertainties). Different void profiles are shown in different colours, as indicated in the legend. The dashed lines show the same models neglecting GR (Equation~\ref{z_contributions}). The observations shown here are indicated using a symbol in the $D_{\mathrm{H}}$ column of Table~\ref{BAO_sample_table}. The size of each point and the thickness of its associated error bar have been scaled based on the year of publication, as indicated towards the lower left. The redshifts shown here have been dithered slightly at $z = 0.32$, 0.51, 1.48, and 2.33 to make the points more clearly visible. The point at $\left(z = 2.33, \alpha_\parallel = 1.0134 \right)$ with an overlaid white cross is a reanalysis of both the more recent and older observations at a similar $z$, so we show the latter with grey error bars.}
    \label{D_H_graph}
\end{figure*}

We next present results on $D_{\mathrm{H}}$, which is inversely related to the redshift depth $\Delta z$ of the BAO ruler (Equation~\ref{Delta_z_BAO}). Our results are shown in Figure~\ref{D_H_graph} in terms of the parameter $\alpha_\parallel$ (Equation~\ref{alpha_parallel}). There is again a small uncertainty in the $\Lambda$CDM expectation because the CMB observations are not perfect. The void models predict that $\alpha_\parallel$ deviates from unity much less than does $\alpha_\perp$. This is because the extra redshift due to a local void necessarily reduces the comoving distance to any fixed redshift, thereby reducing $D_{\mathrm{c}}$. The situation with $D_{\mathrm{H}}$ is more complicated because it depends on the gradient of the outward peculiar velocity (Section~\ref{Redshift_depth}). This is more apparent if considering the dashed lines in Figure~\ref{D_H_graph}, which omit the GR contribution. Without this, the models actually predict $\alpha_\parallel$ values \emph{above} unity at $z \ga 0.15$. This reversed behaviour arises because the outward peculiar velocity from a local void must start declining at some point, reversing the impact on the redshift gradient. BAO observations are generally so distant that they only probe this declining region. Once the GR contribution is included, the Exponential and Gaussian profiles predict $\alpha_\parallel < 1$ but to a much smaller extent, while the Maxwell-Boltzmann profile predicts $\alpha_\parallel > 1$ across a wide range of redshifts.

Many of the observations in this redshift range do indeed show $\alpha_\parallel > 1$, though these are quite old. More recent measurements scatter on both sides of the model prediction by much larger amounts than any deviations predicted in the void models. Measuring $\Delta z$ appears to be considerably harder than measuring $\Delta \theta$, no doubt because $\Delta z$ requires fairly accurate redshifts while $\Delta \theta$ requires accurate angular positions on the sky, which are vastly easier to obtain. As a result, there are fewer $\alpha_\parallel$ measurements and these are less accurate. Importantly for testing a local solution to the Hubble tension, we could not find published $D_{\mathrm{H}}$ measurements at $z < 0.3$. Even so, it is still possible to test the prediction that deviations from unity should be much smaller for $\alpha_\parallel$ than for $\alpha_\perp$. This is broadly in line with the observations, which scatter around unity rather than systematically going below unity, as do the $\alpha_\perp$ values in Figure~\ref{D_c_graph}. Given observational difficulties and the smaller predicted deviations, we conclude that measurements of $\alpha_\parallel$ are presently unable to distinguish between the local void models considered here and the homogeneous \emph{Planck} cosmology.

So far, we have assumed that there is only a very small 0.2\% uncertainty in $r_{\mathrm{d}}$ based on section~5.4 of \citet{Planck_2020}. However, the Hubble tension has motivated some workers to consider new physics that significantly affects the universe prior to recombination, causing large adjustments to what value of $r_{\mathrm{d}}$ is required to fit the CMB anisotropies \citep[e.g.][and references therein]{Poulin_2019, Poulin_2023}. Since we defined $\alpha_\parallel$ and $\alpha_\perp$ with respect to the standard value of $r_{\mathrm{d}}$, a different value would imply that these $\alpha$ parameters should differ from unity, albeit by a constant factor at all $z \la 1000$. Much of the evidence in Figure~\ref{D_c_graph} for deviations from a void-free model would be lost if such a model predicts a flat horizontal line with arbitrary rather than fixed normalization.

\begin{figure*}
    \includegraphics[width=\textwidth]{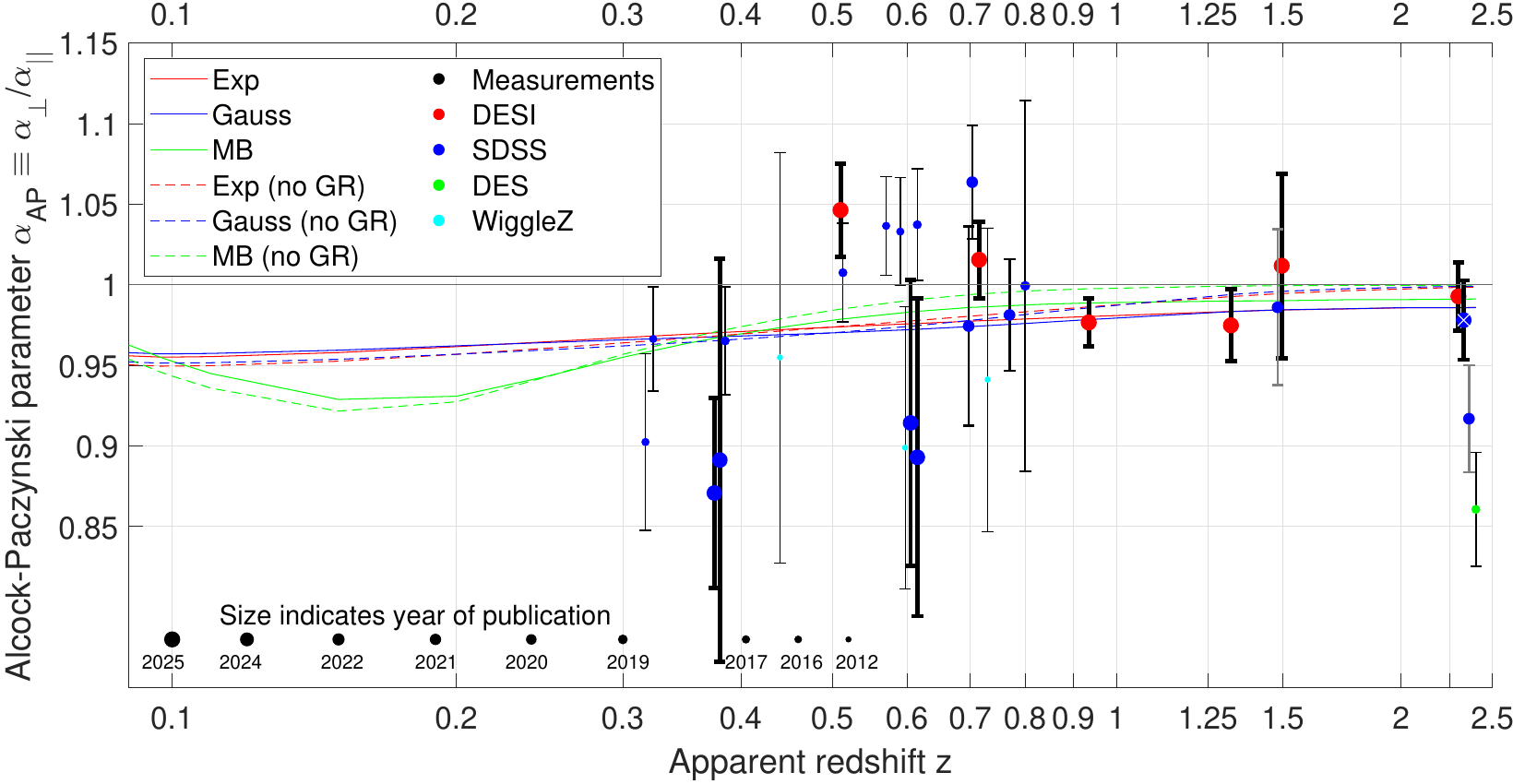}
    \caption{The AP parameter (Equation~\ref{alpha_AP}) as a function of the apparent redshift $z$ according to the model (solid lines) and observations (points with uncertainties). Different void profiles are shown in different colours, as indicated in the legend. The dashed lines show the predictions without GR. The observations shown here are indicated using a symbol in the $\alpha_{_{\mathrm{AP}}}$ column of Table~\ref{BAO_sample_table}. We calculate $\alpha_{_{\mathrm{AP}}}$ in all cases. The size of each point and the thickness of its associated error bar have been scaled based on the year of publication, as indicated towards the lower left. The redshifts shown here have been dithered slightly at $z = 0.32$, 0.51, 1.48, and 2.33 to make the points more clearly visible. The point at $\left(z = 2.33, \alpha_{_{\mathrm{AP}}} = 0.9771 \right)$ with an overlaid white cross is a reanalysis of both the more recent and older observations at a similar $z$, so we show the latter with grey error bars.}
    \label{AP_graph}
\end{figure*}

This motivates us to present results in a way that is not sensitive to the assumed value of $r_{\mathrm{d}}$. We do so using $\alpha_{\mathrm{AP}}$, which cancels out the impact of $r_{\mathrm{d}}$ by taking the ratio of parameters which depend on $r_{\mathrm{d}}$ in the same way (Equation~\ref{alpha_AP}). We use Figure~\ref{AP_graph} to compare the observed and predicted values of $\alpha_{\mathrm{AP}}$. While this is a good idea in principle, we see that taking the ratio of quantities which are already somewhat uncertain further inflates the uncertainties. Moreover, the need for both $D_{\mathrm{c}}$ and $D_{\mathrm{H}}$ measurements from the same survey at the same $z$ limits the number of available measurements, especially at low $z$. It is thus not presently possible to assess if the local void models perform better against the void-free model based on the AP test. In the future, measurements of $\alpha_{\mathrm{AP}}$ at percent-level precision in the critical $z \la 0.3$ regime could be of vital importance to distinguishing if the Hubble tension should be solved through new physics in the early Universe or at late times/locally.

\begin{figure*}
    \includegraphics[width=\textwidth]{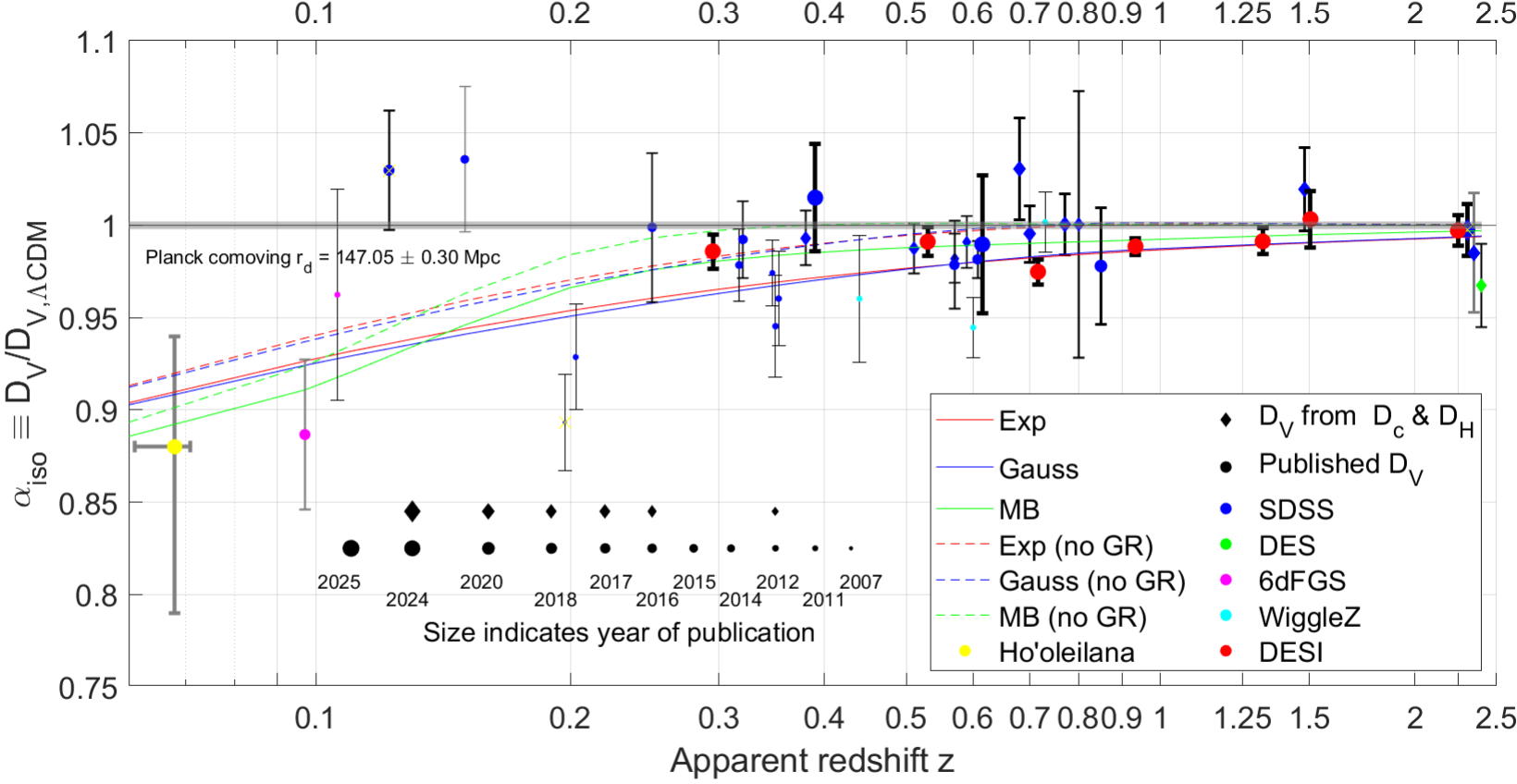}
    \caption{The ratio $\alpha_{\mathrm{iso}}$ (Equation~\ref{alpha_iso}) of the isotropically averaged comoving BAO scale $D_{\mathrm{V}}$ to the $\Lambda$CDM expectation as a function of the apparent redshift $z$, both according to the model (solid lines) and observations (points with uncertainties). Different void profiles are shown in different colours, as indicated in the legend. The dashed lines show these models without GR (Equation~\ref{z_contributions}). The observations shown here are indicated using a symbol in the $D_{\mathrm{V}}$ column of Table~\ref{BAO_sample_table}. The size of each point and the thickness of its associated error bar have been scaled based on the year of publication, as indicated towards the lower left. Diamond symbols indicate that we calculated $D_{\mathrm{V}}$ from the published $D_{\mathrm{c}}$ and $D_{\mathrm{H}}$. The redshifts shown here have been dithered slightly at $z = 0.2$, 0.51, 1.48, and 2.33 to make the points more clearly visible. The points at $\left(z = 0.122, \alpha_{\mathrm{iso}} = 1.0298 \right)$, $\left(z = 0.2, \alpha_{\mathrm{iso}} = 0.8932 \right)$, and $\left(z = 2.33, \alpha_{\mathrm{iso}} = 0.9974 \right)$ with an overlaid cross are reanalyses of both the more recent and older observations at a similar $z$, so we show the latter with grey error bars. The older data used in the point at $z = 0.2$ are not shown here. The lowest redshift point at $z = 0.068$ is shown with grey uncertainties as it represents a single structure (see the text).}
    \label{D_V_graph}
\end{figure*}

Measurement uncertainties can be reduced by averaging different measurements rather than taking their ratio. For this purpose, we consider the isotropically averaged BAO distance $D_{\mathrm{V}}$, which is essentially the geometric mean of $D_{\mathrm{c}}$ and $D_{\mathrm{H}}$, albeit with the former counted twice (Equation~\ref{D_V}). We present our results in Figure~\ref{D_V_graph} in terms of the parameter $\alpha_{\mathrm{iso}}$, which measures deviations from the $\Lambda$CDM expectation (Equation~\ref{alpha_iso}). If $D_{\mathrm{c}}$ and $D_{\mathrm{H}}$ are measured independently and have the same fractional uncertainty, we would expect the fractional uncertainty on $D_{\mathrm{V}}$ to be smaller by a factor of $\sqrt{5/9}$, so we might expect typical uncertainties to be smaller by about 25\% (Equation~\ref{sigma_D_V}).

It is immediately apparent from Figure~\ref{D_V_graph} that not only are the measurements indeed much more accurate, there are also many more of them. This is because $D_{\mathrm{V}}$ can be found by fitting an ellipse to the galaxy/quasar correlation function plotted against both angular separation and redshift depth. In situations where the data are too noisy to permit measuring either the width or the height of this ellipse, it might still be possible to obtain a good estimate of its area. This is why many of the BAO measurements in our compilation only give $D_{\mathrm{V}}$, especially at very low $z$ (Table~\ref{BAO_sample_table}). For all these reasons, the $\alpha_{\mathrm{iso}}$ parameter should provide the clearest indication of the viability of the local void solution to the Hubble tension.

The results in Figure~\ref{D_V_graph} are clearly much more in line with any of the considered void models than with the void-free model. There appear to be two points at $z = 0.12 - 0.15$ that are not in line with this trend, but the uncertainty of the $z = 0.15$ point is shown in grey to indicate that it relies on data that was later reused in the $z = 0.122$ point (the same applies to the point at $z = 0.097$). As a result, the only obvious discrepancy from the local void models is the single point at $z = 0.122$, though the discrepancy is still only $2.9\sigma - 3.1\sigma$ depending on the density profile. Even with the correct model, it is plausible at the 10\% level that there should be a point with this level of tension out of the 42 available $D_{\mathrm{V}}$ measurements, even before considering other issues like systematic errors or non-Gaussian tails. Still, it is clear that this particular point fits much better ($0.95\sigma$ tension) with the void-free model.

Given the importance of data at low $z$, we briefly mention a somewhat uncertain BAO measurement at $z = 0.07$ from an individual BAO called Ho'oleilana \citep{Tully_2023_BAO}. While the BAO scale usually has to be found through careful statistical analysis of many galaxies, a bump in the power spectrum at a particular scale makes it more likely that a large structure of galaxies will have this scale rather than a slightly smaller or larger scale. As a result, the size of the galaxy ``bubble'' identified by those authors can be used as a low precision constraint on the BAO scale at its redshift. Given the difficulties, the authors were inevitably only able to constrain $\alpha_{\mathrm{iso}}$. Comparison with mock data indicates that typically it would have an uncertainty of 0.14, but we are fortunate to be living near a particularly strong and thus clearly defined individual BAO, reducing the uncertainties somewhat. As a result, observations of Ho'oleilana imply that $\alpha_{\mathrm{iso}} = 0.88^{+0.06}_{-0.09}$ \citep{Tully_2023_BAO}. This is in excellent ($<1\sigma$) agreement with any of the considered void models, but $2\sigma$ below the expectation of unity in the void-free model. While one should not read too much into an individual association of galaxies, Ho'oleilana is one of several BAO observations at low $z$ that are more in line with the considered void models. This includes the lowest $z$ measurement of $D_{\mathrm{c}}$ shown in Figure~\ref{D_c_graph} and three DES measurements around $z = 0.8$ shown there in green, all of which lack isotropic counterparts on Figure~\ref{D_V_graph}.

\subsection{Overall goodness of fit}
\label{chi_sq_analysis}

We construct an approximate $\chi^2$ statistic to assess the overall performance of each model against the observations listed in Table~\ref{BAO_sample_table}. Because of the severe complications associated with estimating covariances, we simply assume that these observations are all independent. We try to avoid double-counting data by not considering the points shown there in grey, which are based on data that was later reanalysed and used as part of a different BAO measurement. We also exclude the somewhat speculative data point based on Ho'oleilana. In some cases, results are available from the same study with coarser or finer redshift bins. To avoid crowding the figures, we plot the former, but we use the latter when calculating $\chi^2$ to benefit from a larger number of data points and a wider redshift range.

When considering the parameters $\alpha_\perp$, $\alpha_\parallel$, and $\alpha_{\mathrm{iso}}$, we also need to make an allowance for uncertainty in the \emph{Planck} value of $r_{\mathrm{d}}$ \citep[$147.05 \pm 0.30$~cMpc; see section~5.4 of][]{Planck_2020}. For this, we combine in quadrature the fractional uncertainty in the \emph{Planck} $r_{\mathrm{d}}$ with that on the corresponding $\alpha$ parameter due solely to observational uncertainties in the related BAO observable. Omitting subscripts indicating the particular BAO observable for clarity, the uncertainty on any $\alpha$ parameter related to the BAO distance measure $D$ is thus
\begin{eqnarray}
    \sigma \left( \alpha \right) ~=~ \sqrt{\left[ \sigma \left( \frac{D}{r_{\mathrm{d}}} \right) \div \left. \frac{D}{r_{\mathrm{d}}} \right|_{\mathrm{fid}} \right]^2 \, + \, \left[ \widetilde{\sigma} \left( r_{\mathrm{d}} \right) \alpha_{\mathrm{model}} \right]^2} \, .
    \label{sigma_alpha}
\end{eqnarray}
The idea is that the $\Lambda$CDM prediction of $\alpha \equiv 1$ has an associated uncertainty of $\widetilde{\sigma} \left( r_{\mathrm{d}} \right)$, so in a different model which predicts $\alpha = \alpha_{\mathrm{model}}$, the corresponding uncertainty is $\widetilde{\sigma} \left( r_{\mathrm{d}} \right) \alpha_{\mathrm{model}}$. Combining this with the uncertainty in the BAO observable $D/r_{\mathrm{d}}$, we obtain a slightly inflated $\sigma \left( \alpha \right)$, which we then use as the error budget when calculating the contribution to $\chi^2$.

We do not apply the above procedure to the AP parameter because it is defined so that $r_{\mathrm{d}}$ cancels out (Equation~\ref{alpha_AP}) -- only the observational uncertainty of $\alpha_{_{\mathrm{AP}}}$ is relevant. Since $\alpha_{_{\mathrm{AP}}}$ is almost never available directly, we work it out from the published $D_{\mathrm{c}}/r_{\mathrm{d}}$ and $D_{\mathrm{H}}/r_{\mathrm{d}}$ assuming uncorrelated uncertainties.
\begin{eqnarray}
    \widetilde{\sigma} \left( \alpha_{\mathrm{AP}} \right) ~=~ \sqrt{\widetilde{\sigma}^2 \left( \frac{D_{\mathrm{c}}}{r_{\mathrm{d}}} \right) \, + \, \widetilde{\sigma}^2 \left( \frac{D_{\mathrm{H}}}{r_{\mathrm{d}}} \right)} \, .
    \label{sigma_alpha_AP}
\end{eqnarray}

The $\chi^2$ values obtained in this way are shown in Table~\ref{chi_sq_table}. Different rows show different models, while different columns show different BAO observables and combinations thereof. The number of data points ($N$) is indicated in the bottom row. The numbers outside brackets give the $\chi^2$, while the bracketed numbers immediately below them show the likelihood of a higher $\chi^2$ for $N$ degrees of freedom, which we express as an equivalent number of standard deviations for a single Gaussian random variable \citep[the relation between this `$Z$-score' and the probability is shown in figure~2 of][]{Asencio_2023}. Since BAO observations were not used in fixing the parameters of either the \emph{Planck} cosmology or the local void models considered by \citetalias{Haslbauer_2020}, the number of degrees of freedom is the same as the number of data points, all of which can be used to test the \emph{a priori} model predictions.

\begin{table}
    \centering
    \caption{The numbers outside brackets show the $\chi^2$ of the homogeneous \emph{Planck} cosmology and our local void models (different rows) against different BAO observables (different columns). The last column combines $D_{\mathrm{V}}$ measurements with $D_{\mathrm{c}}$ from studies that only report $D_{\mathrm{c}}$. Results are based on the observations listed in Table~\ref{BAO_sample_table}, which we assume are all independent. We only consider the points shown there in black to avoid double-counting (see the text). The numbers in brackets show the $\chi^2$ value as an equivalent level of tension for a single Gaussian variable, assuming the number of data points in the final row is the number of degrees of freedom. This is justified because none of the models considered here were calibrated using BAO data, which therefore serve as an independent test in all cases.}
    \begin{tabular}{lccccc}
        \hline
        Void & \multicolumn{4}{c}{\raisebox{-2pt}{BAO observable used to calculate $\chi^2$}} & $\alpha_{\mathrm{iso}}$ \& \\
        profile & $\alpha_\perp$ ($D_{\mathrm{c}}$) & $\alpha_\parallel$ ($D_{\mathrm{H}}$) & $\alpha_{_{\mathrm{AP}}}$ & $\alpha_{\mathrm{iso}}$ ($D_{\mathrm{V}}$) & only $\alpha_\perp$ \\ [3pt] \hline
        $\Lambda$CDM & 47.14 & 36.91 & 50.56 & 75.75 & 93.03 \\
        (no void) & ($1.64\sigma$) & ($1.44\sigma$) & ($2.66\sigma$) & ($3.27\sigma$) & ($3.79\sigma$) \\ [3pt]
        \multirow{2}{*}{Exponential} & 46.07 & 34.96 & 50.15 & 50.31 & 55.85 \\
        & ($1.55\sigma$) & ($1.26\sigma$) & ($2.62\sigma$) & ($1.35\sigma$) & ($1.19\sigma$) \\ [3pt]
        \multirow{2}{*}{Gaussian} & 48.74 & 36.34 & 52.72 & 51.25 & 56.76 \\
        & ($1.77\sigma$) & ($1.39\sigma$) & ($2.84\sigma$) & ($1.42\sigma$) & ($1.26\sigma$) \\ [3pt]
        Maxwell- & 31.05 & 35.90 & 48.22 & 47.25 & 54.74 \\
        Boltzmann & ($0.38\sigma$) & ($1.35\sigma$) & ($2.46\sigma$) & ($1.11\sigma$) & ($1.11\sigma$) \\ \hline
        Data points & 36 & 29 & 29 & 42 & 49 \\ \hline
    \end{tabular}
    \label{chi_sq_table}
\end{table}

Our results indicate that compared to the void models, the homogeneous model generally provides a poorer fit to the BAO data. The tension of $\la 1.8\sigma$ is however not significant for $D_{\mathrm{c}}$ and $D_{\mathrm{H}}$, and in any case the void models only slightly reduce it. Taking the ratio of $D_{\mathrm{c}}$ and $D_{\mathrm{H}}$ via the AP parameter (Equation~\ref{alpha_AP}) eliminates dependence on $r_{\mathrm{d}}$, but it also inflates the uncertainties. All models are in moderate ($2.5\sigma - 2.8\sigma$) tension with the observed values of $\alpha_{_{\mathrm{AP}}}$. The greatest tension arises with the Gaussian void model, but the other void models slightly outperform $\Lambda$CDM.

Since we work out $\alpha_{_{\mathrm{AP}}}$ in all cases assuming independent errors on $D_{\mathrm{c}}$ and $D_{\mathrm{H}}$, the moderate tension faced by all considered models may indicate that the uncertainties on $D_{\mathrm{c}}$ and $D_{\mathrm{H}}$ are in fact correlated to some extent. We do not take this into account because the correlation is often not available for earlier studies, and our aim is to have a consistent treatment of all available BAO measurements. However, we can gain some idea of how this might impact our results from the final column of table~4 in \citet{DESI_2025}, which shows that uncertainties on $D_{\mathrm{c}}$ and $D_{\mathrm{H}}$ are typically anti-correlated. Since $\alpha_{_{\mathrm{AP}}} \propto D_{\mathrm{c}}/D_{\mathrm{H}}$ (Equation~\ref{alpha_AP}), treating their uncertainties as uncorrelated leads to an underestimated uncertainty on their ratio, and thus on $\alpha_{_{\mathrm{AP}}}$. Bearing this in mind, our estimated $<3\sigma$ tension in $\alpha_{_{\mathrm{AP}}}$ for all considered models does not seem particularly problematic.

If $D_{\mathrm{c}}$ and $D_{\mathrm{H}}$ typically have anti-correlated errors \citep[as they do in DESI~DR2;][]{DESI_2025}, then we might get more accurate results not by taking the ratio of $D_{\mathrm{c}}$ and $D_{\mathrm{H}}$ but instead by taking their product, which is essentially what $D_{\mathrm{V}}$ represents (Equation~\ref{D_V}). There are also more available measurements of $\alpha_{\mathrm{iso}}$ than for any other $\alpha$ parameter. This includes several measurements at very low $z$, which is particularly crucial when testing the local void scenario.

The second-last column of Table~\ref{chi_sq_table} shows that the void-free model faces $3.3\sigma$ tension with respect to $\alpha_{\mathrm{iso}}$, but this drops to only $1.1\sigma - 1.4\sigma$ when considering the void models, depending on the adopted density profile. This is because $\chi^2$ decreases from 76 with no void to $47-51$ in the void models, which are thus able to reduce $\chi^2$ by about 25. \emph{A direct comparison between the homogeneous model and any of the considered void models prefers the latter by a likelihood ratio similar to that between a model which is $5\sigma$ discrepant from an observation with Gaussian errors and a different model that exactly matches the observation.} We were generally able to use published $D_{\mathrm{V}}$ measurements and their uncertainties, thereby avoiding inaccuracies associated with assuming uncorrelated errors on $D_{\mathrm{c}}$ and $D_{\mathrm{H}}$ (this is indicated by most of the symbols in the $D_{\mathrm{V}}$ column of Table~\ref{BAO_sample_table} being circles rather than diamonds, especially at low $z$). In the remaining cases, our assumption is likely to overestimate the error on $D_{\mathrm{V}}$ for the same reason that it underestimates the error on $\alpha_{\mathrm{AP}}$ (see above). Therefore, the tensions faced by all models from the observed values of $\alpha_{\mathrm{iso}}$ are likely to be slightly higher than reported here. This would not be problematic for the void models because we estimate an overall tension of only $1.1\sigma - 1.4\sigma$, but raising the $3.3\sigma$ tension for the void-free model would make it even less plausible. Moreover, these tensions consider only 42 measurements of $D_{\mathrm{V}}$, which misses a few studies that report only $D_{\mathrm{c}}$ and thus cannot yield information on $D_{\mathrm{V}}$ (Table~\ref{BAO_sample_table}). Including these 7 studies to get a more complete picture, we can get a total of 49 points, the results for which are shown in the final column of Table~\ref{chi_sq_table}. The considerable $3.8\sigma$ tension with $\Lambda$CDM decreases to only $1.1\sigma - 1.3\sigma$ in the void models. Our results therefore support the local void solution to the Hubble tension.

Comparing the different void models, the Maxwell-Boltzmann profile generally achieves a better fit to the data compared to the other void profiles, though its performance is very slightly worse than that of the Exponential profile when considering $D_{\mathrm{H}}$. In the case of $D_{\mathrm{V}}$ (which we consider most informative due to the larger amount of data and smaller uncertainties), differences between the different void profiles are very minor, and certainly much less significant than the difference between any of the void models and the homogeneous model. The only major difference between the Maxwell-Boltzmann profile and the other void profiles arises when considering $D_{\mathrm{c}}$, in which case it is able to reduce $\chi^2$ by $15-18$ compared to the other profiles. This indicates some preference for the Maxwell-Boltzmann profile, but there are also strong arguments against it based on the bulk flow curve \citep{Mazurenko_2024}. Comparing the Exponential and Gaussian profiles, the former consistently achieves a lower $\chi^2$ by $0.9 - 2.7$, a negligible difference. Our results thus shed little light on which of these should be preferred.

\subsection{Comparison with DESI DR2}
\label{DESI_comparison}

DESI is the largest BAO survey conducted so far \citep{DESI_2016, DESI_2022}. Its recent DR2 \citep{DESI_2025} used data on $>14$~million galaxies and $>1$~million quasars \citep{DESI_2025_quasars} to quantify the BAO angular scale and redshift depth over the range $z = 0.3 - 2.3$. While DESI is by no means the only BAO survey and several other studies add crucial information at $z < 0.3$ (Table~\ref{BAO_sample_table}), it is interesting to compare $\Lambda$CDM and the \citetalias{Haslbauer_2020} void models with DESI~DR2, which provides 7 measurements of $D_{\mathrm{V}}/r_{\mathrm{d}}$ \citep[table~4 of][]{DESI_2025}. These give a total $\chi^2$ and Gaussian equivalent tension of 23.27 ($3.17\sigma$) for $\Lambda$CDM, which the void models reduce to 10.94 ($1.47\sigma$), 12.26 ($1.68\sigma$), and 6.24 ($0.66\sigma$) for the Exponential, Gaussian, and Maxwell-Boltzmann profile, respectively. The isotropic BAO distances from DESI~DR2 are in considerable tension with the void-free model and much more in line with any of the void models.

It is possible to do a more detailed comparison with the DESI~DR2 measurements by jointly considering $D_{\mathrm{c}}/r_{\mathrm{d}}$ and $D_{\mathrm{H}}/r_{\mathrm{d}}$, removing the need to consider combinations like $D_{\mathrm{V}}/r_{\mathrm{d}}$ or $\alpha_{\mathrm{AP}}$. In this case, it is important to take into account the published correlation coefficient $\rho_{\mathrm{c,H}}$ between $D_{\mathrm{c}}/r_{\mathrm{d}}$ and $D_{\mathrm{H}}/r_{\mathrm{d}}$ at each $z$ \citep[see the final column of table~4 in][]{DESI_2025}. By definition, the covariance between these observables is
\begin{eqnarray}
    \Cov \left( \frac{D_{\mathrm{c}}}{r_{\mathrm{d}}}, \frac{D_{\mathrm{H}}}{r_{\mathrm{d}}} \right) ~\equiv~ \sigma \left( \frac{D_{\mathrm{c}}}{r_{\mathrm{d}}} \right) \sigma \left( \frac{D_{\mathrm{H}}}{r_{\mathrm{d}}} \right) \rho_{\mathrm{c,H}} \, ,
\end{eqnarray}
where $\Cov \left( x, y \right)$ denotes the covariance between any two quantities $x$ and $y$ with correlation coefficient $\rho_{x,y}$ or $\rho \left( x, y \right)$. Including an allowance for uncertainty in the \emph{Planck} $r_{\mathrm{d}}$ (Equation~\ref{sigma_alpha}) and bearing in mind that it leads to a systematic error in both $\alpha_\perp$ and $\alpha_\parallel$, we get that their covariance is
\begin{eqnarray}
    \Cov \left( \alpha_\perp, \alpha_\parallel \right) ~=~ \frac{\Cov \left( \frac{D_{\mathrm{c}}}{r_{\mathrm{d}}}, \frac{D_{\mathrm{H}}}{r_{\mathrm{d}}} \right)}{\left. \frac{D_{\mathrm{c}}}{r_{\mathrm{d}}} \right|_{\mathrm{fid}} \left. \frac{D_{\mathrm{H}}}{r_{\mathrm{d}}} \right|_{\mathrm{fid}}} \, + \, \widetilde{\sigma}^2 \left( r_{\mathrm{d}} \right) \alpha_{\perp}^{\mathrm{model}} \alpha_{\parallel}^{\mathrm{model}} \, .
\end{eqnarray}
We can then work out the implied correlation coefficient.
\begin{eqnarray}
    \rho \left( \alpha_\perp, \alpha_\parallel \right) ~\equiv~ \frac{\Cov \left( \alpha_\perp, \alpha_\parallel \right)}{\sigma \left( \alpha_\perp \right) \sigma \left( \alpha_\parallel \right)} \, ,
\end{eqnarray}
with the uncertainties in the denominator found using Equation~\ref{sigma_alpha}.

Having obtained the uncertainties on $\alpha_\perp$ and $\alpha_\parallel$ and their mutual correlation coefficient $\rho \left( \alpha_\perp, \alpha_\parallel \right)$, we can find the combined $\chi^2$ with respect to the model predictions. We first define
\begin{eqnarray}
    \chi_\perp ~\equiv~ \frac{\alpha_\perp - \alpha_{\perp}^{\mathrm{model}}}{\sigma \left( \alpha_\perp \right)} \, ,
\end{eqnarray}
with $\chi_\parallel$ defined analogously. We then apply equation~66 of \citetalias{Haslbauer_2020} to find the total $\chi^2$ including the correlation term.
\begin{eqnarray}
    \chi^2 ~=~ \frac{1}{2} \left[ \frac{\left( \chi_\perp + \chi_\parallel \right)^2}{1 + \rho \left( \alpha_\perp, \alpha_\parallel \right)} + \frac{\left( \chi_\perp - \chi_\parallel \right)^2}{1 - \rho \left( \alpha_\perp, \alpha_\parallel \right)} \right] \, .
    \label{chi_sq_2D}
\end{eqnarray}
This reduces to the standard expression $\chi^2 = \chi_\perp^2 + \chi_\parallel^2$ if $\rho \left( \alpha_\perp, \alpha_\parallel \right) = 0$.

We find $\chi^2$ separately for each of the six redshifts at which both $D_{\mathrm{c}}/r_{\mathrm{d}}$ and $D_{\mathrm{H}}/r_{\mathrm{d}}$ are reported in table~4 of \citet{DESI_2025}. In each case, we work out the Gaussian equivalent tension assuming 2 degrees of freedom, in which case the likelihood of a higher $\chi^2$ with the correct model is $\exp \left( - \chi^2/2 \right)$. We also find the total $\chi^2$ across these six redshift bins and a seventh bin at $z = 0.295$ where only $D_{\mathrm{V}}/r_{\mathrm{d}}$ is available, removing the need to consider correlated uncertainties. The total $\chi^2$ therefore corresponds to 13 degrees of freedom.

\begin{table}
    \centering
    \caption{The numbers outside brackets show the $\chi^2$ of the homogeneous \emph{Planck} cosmology and our local void models (different columns) against the BAO observables from DESI~DR2 \citep{DESI_2025} at each $z$ (different rows). Except for the lowest redshift of $z = 0.295$ where only $D_{\mathrm{V}}/r_{\mathrm{d}}$ is available, the $\chi^2$ considers both $D_{\mathrm{c}}/r_{\mathrm{d}}$ and $D_{\mathrm{H}}/r_{\mathrm{d}}$, allowing for their correlated uncertainties (Equation~\ref{chi_sq_2D}) and uncertainty in the \emph{Planck} $r_{\mathrm{d}}$ (Equation~\ref{sigma_alpha}). The numbers in brackets show the $\chi^2$ as an equivalent level of tension for a single Gaussian variable, assuming 2 degrees of freedom (1 at $z = 0.295$). The final two rows show the total $\chi^2$ and Gaussian equivalent tension for 13 degrees of freedom.}
    \begin{tabular}{lcccc}
        \hline
        & \multicolumn{4}{c}{Void profile} \\
        & $\Lambda$CDM & & & Maxwell-\\
        $z$ & (no void) & Exponential & Gaussian & Boltzmann \\ [3pt] \hline
        \multirow{2}{*}{0.295} & 2.19 & 4.87 & 6.02 & 0.35 \\
        & ($1.48\sigma$) & ($2.21\sigma$) & ($2.45\sigma$) & ($0.59\sigma$) \\ [3pt]
        \multirow{2}{*}{0.51} & 4.36 & 9.41 & 10.21 & 6.07 \\
        & ($1.58\sigma$) & ($2.61\sigma$) & ($2.74\sigma$) & ($1.98\sigma$) \\ [3pt]
        \multirow{2}{*}{0.706} & 13.83 & 3.95 & 4.69 & 6.64 \\
        & ($3.29\sigma$) & ($1.48\sigma$) & ($1.66\sigma$) & ($2.10\sigma$) \\ [3pt]
        \multirow{2}{*}{0.934} & 7.94 & 0.39 & 0.19 & 1.11 \\
        & ($2.35\sigma$) & ($0.22\sigma$) & ($0.11\sigma$) & ($0.56\sigma$) \\ [3pt]
        \multirow{2}{*}{1.321} & 2.22 & 0.28 & 0.24 & 0.52 \\
        & ($0.97\sigma$) & ($0.16\sigma$) & ($0.14\sigma$) & ($0.29\sigma$) \\ [3pt]
        \multirow{2}{*}{1.484} & 0.11 & 0.96 & 0.93 & 0.47 \\
        & ($0.06\sigma$) & ($0.50\sigma$) & ($0.48\sigma$) & ($0.27\sigma$) \\ [3pt]
        \multirow{2}{*}{2.33} & 0.14 & 0.21 & 0.18 & 0.01 \\
        & ($0.08\sigma$) & ($0.13\sigma$) & ($0.11\sigma$) & ($0.00\sigma$) \\ \hline
        Combined & 30.78 & 20.07 & 22.45 & 15.16 \\
        tension & ($2.91\sigma$) & ($1.68\sigma$) & ($1.97\sigma$) & ($1.04\sigma$) \\ \hline
    \end{tabular}
    \label{chi_sq_table_DESI_DR2}
\end{table}

Our results are shown in Table~\ref{chi_sq_table_DESI_DR2}, with the bracketed numbers indicating the Gaussian equivalent tension for the corresponding number of degrees of freedom. Working inwards from high $z$ where all models give similar predictions, we see that there is very little tension down to $z = 1$ in all cases. At lower $z$, the data prefer the void models, with the results at $z = 0.706$ and 0.934 showing a particularly strong preference. The only redshift which appreciably disfavours the void models is $z = 0.51$, but here $\Lambda$CDM is only favoured by $\Delta \chi^2 = 2-6$, which is not particularly compelling given this is the most favourable DESI~DR2 data point for $\Lambda$CDM with the \citet{Planck_2020} parameters. The lowest redshift BAO measurement from DESI~DR2 is at $z = 0.295$, though this only covers $D_{\mathrm{V}}/r_{\mathrm{d}}$. The isotropic BAO results here mildly disfavour the Exponential and Gaussian void models over $\Lambda$CDM, though the Maxwell-Boltzmann void model does slightly better still.

The overall picture is one of good agreement with all models at high $z$ where they agree with each other, but where they diverge at $z \la 1$, the void models are preferred by DESI~DR2. $\Lambda$CDM faces an overall tension of $2.9\sigma$, but this drops to only $1.0\sigma - 2.0\sigma$ in the void models. The total $\chi^2$ with respect to the 13 available BAO measurements favours the void models by $\Delta \chi^2 = 8.3 - 15.6$, depending on the considered density profile. Taking the Exponential profile which gives an intermediate $\Delta \chi^2 = 10.7$, we find that the local void scenario is $210\times$ more likely than the homogeneous \emph{Planck} cosmology based on DESI~DR2 alone \citep[table~4 of][]{DESI_2025}. Thus, the preference for the void models apparent in our main analysis (Table~\ref{chi_sq_table}) is also apparent in DESI~DR2, albeit at lower significance. This is to be expected for a genuine signal given our main analysis considers a much larger number of BAO measurements, many of which extend to the particularly important $z \la 0.3$ regime. Neither our main analysis nor our comparison with DESI~DR2 sheds much light on which of the void density profiles is to be preferred, though they all provide a better fit than $\Lambda$CDM.

\section{Discussion}
\label{Discussion}

Following equation~4 of \citet{Mazurenko_2025} and assuming an updated $H_0^{\mathrm{Planck}} = 67.64 \pm 0.52$~km/s/Mpc \citep{Tristram_2024} and local $cz' = 73.17 \pm 0.86$~km/s/Mpc \citep{Breuval_2024}, the Hubble tension can be summarized as the statement that locally,
\begin{eqnarray}
    cz' ~=~ \left( 1.082 \pm 0.014 \right) H_0^{\mathrm{Planck}} \, .
\end{eqnarray}
There is now increasingly strong evidence for the Hubble tension \citep[Section~\ref{Introduction}; see also][]{Uddin_2024, Jensen_2025}. Given the excellent fit to the CMB anisotropies in $\Lambda$CDM with the \emph{Planck} parameters \citep{Planck_2020, Tristram_2024, Calabrese_2025} and that the Hubble tension is largely a mismatch between the local $cz'$ and the present $\dot{a}$ estimated in other ways \citep{Perivolaropoulos_2024}, the Hubble tension might indicate new physics at late times. This could be due to a local inhomogeneity, or it could indicate an adjustment to the background expansion history at late times. The latter possibility would allow us to continue making the usual assumption that $cz' = \dot{a}$, which must be the case if the Universe is homogeneous on the relevant scales \citep[see equation~3 of][]{Mazurenko_2025}. In either scenario, the expansion history would have followed the \emph{Planck} cosmology for the vast majority of cosmic history. However, there would be an unexpected effect at low $z$ that reduces the distance to any fixed $z$ by about 10\%. Roughly speaking, we would expect this to reduce the BAO distance measures by about 10\% compared to the predictions of the homogeneous \emph{Planck} cosmology. Our results in Figure~\ref{D_V_graph} hint at just this kind of deviation in $D_{\mathrm{V}}$, for which the quality and quantity of observations are better than with other common ways of presenting BAO results. While this could be a coincidence, it is at the least very interesting that BAO distance measures seem to deviate from expectations in the homogeneous \emph{Planck} cosmology in just the way that might be expected if the Hubble tension is genuine and arises at late times.

Our focus in this contribution is on the local void scenario \citepalias{Haslbauer_2020}. A local underdensity was originally proposed long before the Hubble tension based on galaxy number counts in the optical \citep{Maddox_1990, Shanks_1990} and especially in the near-infrared \citep{Keenan_2013}. Based on the results of the latter study, \citetalias{Haslbauer_2020} estimated in their equation~5 that the apparent local expansion rate ($cz'$) should exceed $H_0^{\mathrm{Planck}}$ by $11 \pm 2\%$. \citetalias{Haslbauer_2020} then constructed a semi-analytic model for the formation of a local void from a small initial density fluctuation at high redshift. The void parameters were mostly constrained using the local $cz'$ and the density profile of the KBC void, though the peculiar velocity of the Local Group and strong lensing time delays also played a role. Importantly, \citetalias{Haslbauer_2020} did not consider BAO data when fitting their model parameters. Since we use the best-fitting void parameters from tables~4 and C1 of \citetalias{Haslbauer_2020}, our predictions for the BAO observables can be considered parameter-free \emph{a priori} predictions of the local void scenario. In this respect, it is very encouraging for this scenario that depending on the void density profile and assuming independence of the BAO observations, the isotropic distances prefer the void model by $\Delta \chi^2 = 24.5-28.5$ in terms of $D_{\mathrm{V}}$ (Table~\ref{chi_sq_table}), implying the void models are more likely than the void-free model by a factor of $\exp \left( \Delta \chi^2/2 \right) > 10^5$. In terms of the Akaike Information Criterion \citep[AIC;][]{Akaike_1974} or the Bayesian Information Criterion \citep[BIC;][]{Schwarz_1978}, there is no `Occam penalty' \citep[section 3.1 of][]{Haslbauer_2024} associated with the higher complexity of the void models because their parameters were fixed in \citetalias{Haslbauer_2020} without considering BAO datasets. As a result, both the AIC and the BIC prefer the void scenario over the homogeneous model by the likelihood ratio of $\exp \left( \Delta \chi^2/2 \right) > 10^5$.

Given that the number of data points is 42, the overall tension with the isotropic BAO data drops from $3.3\sigma$ without a void to $1.1\sigma - 1.4\sigma$ in the void model, depending on the assumed initial density profile. The preference is weaker with other distance measures like $D_{\mathrm{c}}$ and $D_{\mathrm{H}}$, though the number of available observations and their quality are both lower in these cases. Even so, the lowest $\chi^2$ always arises with one or other void model, never with the void-free model. While it would be helpful to cancel out the dependence on $r_{\mathrm{d}}$ through use of the AP parameter, it appears that BAO results are presently not well suited to such an analysis, though this could be promising longer term \citep[see the top right panel of figure~6 in][]{DESI_2025}.

Although the trend for BAO observables to deviate from expectations in the homogeneous \emph{Planck} cosmology particularly at low $z$ supports the local void scenario, this does not uniquely imply its validity. The results might instead be due to a homogeneous cosmology whose expansion history deviates from the \emph{Planck} cosmology at late times. Indeed, deviations of the BAO observables from those predicted in the \emph{Planck} cosmology have been interpreted as due to evolution of the dark energy density \citep{Giare_2024, Wang_2024_BAO}, at least if its equation of state $w$ is parametrized using the common form $w = w_0 \, + \, w_a \left( 1 - a \right)$ using the two free parameters $w_0$ and $w_a$ \citep{Chevallier_2001, Linder_2003}. The preferred region of this parameter space indicates that $w < -1$ for a significant portion of cosmic history \citep{Giare_2025, Keeley_2025}, violating the weak or null energy condition \citep{Sen_2008, Lewis_2025}. An increasing dark energy density is also opposite to the model of \citet{Montani_2025}, which had some success in solving the Hubble tension by postulating decay of dark energy into dark matter, which then dilutes with the cosmic expansion. Models designed to fit the BAO preference for $w < -1$ at intermediate redshifts struggle to solve the Hubble tension \citep{Lopez_2025}. This is presumably because $w < -1$ leads to a higher dark energy density and $H \left( z \right)$ than in the \emph{Planck} cosmology, thereby reducing $D_{\mathrm{A}}^{\mathrm{CMB}}$, the angular diameter distance to the CMB (Equation~\ref{D_c}). Given $D_{\mathrm{A}}^{\mathrm{CMB}}$ is tightly constrained from the CMB angular power spectrum \citep{Tristram_2024}, there must be a compensatory period at low $z$ where $H \left( z \right)$ is \emph{less} than in the \emph{Planck} cosmology, worsening the Hubble tension \citep*{Seyed_2025}.\footnote{This period is presumably responsible for the inference that $w_0 > -1$.} We note that since BAO results over the last twenty years are in good agreement with the local void scenario (Table~\ref{chi_sq_table}) and this assumes a constant dark energy density ($w_0 = -1, w_a = 0$), the preference for dynamical dark energy is unlikely to persist in the void model. Additional observations at low $z$ would be required to distinguish a local void from homogeneous models which modify the background expansion history at late times while maintaining $D_{\mathrm{A}}^{\mathrm{CMB}}$.

In this regard, we note that \citet*{Huang_2025} claimed to find ``strong evidence against homogeneous new physics over $\Lambda$CDM'' based on a compilation of SNe, BAO, and cosmic chronometer (CC) data, with the authors suggesting instead that the solution lies with ``local-scale inhomogeneous new physics disguised as local observational systematics''. Moreover, a slightly modified expansion history with standard growth of structure on $\ga 100$~Mpc scales would not explain the observed size and depth of the KBC void, whose properties are in $6.0\sigma$ tension with $\Lambda$CDM expectations \citepalias{Haslbauer_2020}. Nor would it explain the anomalously fast observed bulk flows \citep{Watkins_2023, Whitford_2023}. Additional evidence for enhanced growth of structure is provided by the high redshift, mass, and collision velocity of the interacting galaxy clusters known as El Gordo \citep{Menanteau_2012, Zhang_2015}, whose properties are incompatible with $\Lambda$CDM at $>5\sigma$ confidence for any plausible collision velocity \citep*{Asencio_2021, Asencio_2023}. Giant arcs and rings of quasar absorbers may also be in tension with $\Lambda$CDM \citep*{Lopez_2022, Lopez_2024}, though a more detailed comparison with $\Lambda$CDM expectations accounting for the biased nature of the tracers suggests little tension \citep{Sawala_2025}. This issue is not important in the galaxy number counts of \citet{Keenan_2013} because their survey covers most ($57 - 75\%$) of the galaxy luminosity function over 90\% of the sky (see their figure~9b).

To some extent, the Hubble tension and the tension between BAO observables and predictions in the homogeneous \emph{Planck} cosmology could be addressed by altering $r_{\mathrm{d}}$, which would require new physics prior to recombination. However, early-time solutions to the Hubble tension face difficulties fitting the CMB anisotropies. Since uncalibrated BAO results tightly constrain $\Omega_{\mathrm{M}}$ to very near the \emph{Planck} value \citep{Lin_2021}, a higher $H_0$ implies a higher matter density as the critical density $\propto H_0^2$, but the observed CMB temperature tightly constrains the radiation energy density regardless of $H_0$. Thus, models with a higher $H_0$ and no new physics at $z \la 1000$ would have a different ratio between the matter and radiation densities at recombination. This would alter the pattern of acoustic oscillations in the CMB power spectrum, requiring careful fine-tuning to recover the excellent fit obtained in the $\Lambda$CDM framework \citep{Calabrese_2025}. There might also be an impact on the epoch of matter-radiation equality \citep[when the universe is $7\times$ younger than at recombination;][]{Banik_2025}, changing the turnover scale in the matter power spectrum \citep{Eisenstein_1998}. However, the observed matter power spectrum favours $H_0 = H_0^{\mathrm{Planck}}$ \citep{Philcox_2022, Zaborowski_2024, Farren_2025}. These and other problems with early-time solutions to the Hubble tension were reviewed in \citet{Vagnozzi_2023}, which focused on seven main issues.

Regardless of the CMB anisotropies, BAO observations at high $z$ impose strong constraints on the extent to which merely altering $r_{\mathrm{d}}$ can account for the BAO anomalies. The highest redshift DESI~DR2 measurement of $D_{\mathrm{V}}/r_{\mathrm{d}}$ at $z = 2.33$ implies that any change should be $\la 1\%$ (see the right-most solid red point on Figure~\ref{D_V_graph}). This is part of a wider problem: the BAO results suggest a trend in $\alpha_\perp$ and $\alpha_{\mathrm{iso}}$ with $z$, not a fixed offset from unity at all $z$ \citep{Mukherjee_2025}. Additional BAO observations at $z \la 0.3$ will be crucial to assessing the significance of these trends (see also Figure~\ref{D_c_graph}).

While our results seem to favour a late-time or local solution to the Hubble tension, it is worth discussing solutions at early times, especially through new physics prior to recombination that may allow the CMB anisotropies to be fit with higher $H_0$. An important consequence of such models is that the expansion history would closely follow the \emph{Planck} cosmology for the vast majority of cosmic history ($z \la 1000$), but with a compressed timeline due to the higher $H_0$. It is possible to test this using the ages of the oldest Galactic stars \citep{Cimatti_2023}. Allowing a modest amount of time between the Big Bang and the formation of the considered stars, the resulting $H_0$ value is shown using a magenta band on figure~3 of \citet{Mazurenko_2025}. It is clear that $H_0$ obtained in this way is much more in line with $H_0^{\mathrm{Planck}}$ than the local $cz'$. This is because the Universe has an age similar to that in the \emph{Planck} cosmology, so reducing the predicted age by almost a Gyr leads to many stellar ages exceeding the revised cosmic age \citep[see also][]{Valcin_2025, Xiang_2025}.

The above result is based on the absolute ages of stars, but it is possible to use instead differential ages in the CC technique \citep{Jimenez_2002, Moresco_2018, Moresco_2020, Moresco_2024}. The chronometers in this case are massive quiescent galaxies that stopped forming stars at early times. Detailed analysis of their spectra gives an idea of their age, which can be compared between populations at different $z$ to get the elapsed time. In this way, it is possible to obtain the slope of the time-redshift relation. Results from this technique in combination with other uncalibrated datasets (which cannot constrain $H_0$ on their own) give $H_0$ values very close to $H_0^{\mathrm{Planck}}$ \citep{Cogato_2024, Guo_2025}. Moreover, the latter authors found that the inferred $H_0$ is $>4\sigma$ below the local $cz'$. It is important to note that CC results require a long time baseline to quantify $\dot{z}$, so they probe the cosmology at redshifts where a local void would have little impact.

These results are part of a broader trend for the Hubble tension to become weaker at high redshift, and not merely due to larger uncertainties \citep{Bousis_2024}. The redshift dependence of the Hubble tension was explored in some detail by \citet{Mazurenko_2025} using the concept of $H_0 \left( z \right)$, the value of $H_0$ that would be inferred by observers using data in a narrow redshift range centred on $z$. Using three methods to estimate how observers might go about extrapolating high-redshift results to obtain the present expansion rate, those authors predicted $H_0 \left( z \right)$ using the local void models considered by \citetalias{Haslbauer_2020}, again restricting to their best-fitting parameters for each of the three density profiles they considered. In this way, \citet{Mazurenko_2025} found quite good agreement with recent observational results \citep{Jia_2023, Jia_2025}. In particular, the void models predict a curve that goes approximately halfway between the observational results of these two studies \citep[see figure~3 of][]{Mazurenko_2025}. An early-time solution to the Hubble tension should lead to a flat $H_0 \left( z \right)$ curve at the level of the local $cz'$. However, a descending trend is detected at $5.6\sigma$ significance \citep*{Jia_2023}, rising to $6.4\sigma$ upon including the DESI Y1 BAO results \citep{Jia_2025}. Their empirical $H_0 \left( z \right)$ curve declines from the local $cz'$ and becomes approximately flat at $H_0^{\mathrm{Planck}}$ for $z \ga 0.6$, suggesting that the background expansion history was following the \emph{Planck} cosmology until fairly recently. However, an important deficiency of their analysis is the lack of distinction between angular BAO measurements and those using the redshift depth (Sections~\ref{Angular_scale} and \ref{Redshift_depth}, respectively). There are some hints that these give discrepant results if a homogeneous cosmology is assumed \citep*{Favale_2024}. We address this shortcoming by directly computing how each type of BAO measurement is affected by a local void. This reveals important differences that may be hard to reconcile purely by changing the background cosmology or the assumed $r_{\mathrm{d}}$.

\section{Conclusions}
\label{Conclusions}

Redshifts in the local Universe increase more rapidly with distance than expected in the $\Lambda$CDM cosmological paradigm if its parameters are calibrated to fit the CMB anisotropies. This Hubble tension might arise from outflows driven by a local supervoid, which interestingly is suggested by source number counts across the whole electromagnetic spectrum \citepalias[see section~1.1 of][and references therein]{Haslbauer_2020}. Those authors constructed a semi-analytic model for the formation of the KBC void out of a small initial density fluctuation at high redshift in a background \emph{Planck} cosmology \citep{Planck_2020}, preserving the success of $\Lambda$CDM in fitting the CMB anisotropies \citep{Calabrese_2025} due to not changing the physics in the early universe or $D_{\mathrm{A}}^{\mathrm{CMB}}$ (Appendix~\ref{CMB_impact_void}). $\Lambda$CDM cannot account for the observed KBC void \citepalias{Haslbauer_2020} and does not permit us to reside within a void large and deep enough to solve the Hubble tension \citep{Wu_2017, Camarena_2018}, even though this is precisely what the KBC void appears to be. This led \citetalias{Haslbauer_2020} to conduct semi-analytic simulations in the MOND framework, thereby enhancing the growth of structure on the $\ga 100$~Mpc scales relevant here.

\citetalias{Haslbauer_2020} obtained best-fitting parameters for three different void density profiles (Exponential, Gaussian, and Maxwell-Boltzmann) by considering the fit to various observables, the main ones being the local redshift gradient and the observed density profile of the KBC void. A wide range of common objections to the local void scenario were addressed in section~5.3 of \citetalias{Haslbauer_2020}, including the argument that the large predicted peculiar velocities require our location within the KBC void to be fine-tuned and would lead to a large unobserved kinematic Sunyaev-Zel'dovich (kSZ) effect \citep*{Sunyaev_1980, Ding_2020, Cai_2025}. Neither argument is a serious challenge to the local void scenario, though they are problematic for much larger and deeper voids. At the other extreme, \citet{Wu_2017} assumed too small a size for the local void, thereby erroneously concluding that it needs to be implausibly empty to solve the Hubble tension \citepalias[see section~5.3.5 of][]{Haslbauer_2020}.

In this contribution, we explore the consequences of the \citetalias{Haslbauer_2020} void models for BAO datasets, which were not considered by \citetalias{Haslbauer_2020}. We do not vary their best-fitting model parameters for each of the three density profiles they considered, thereby obtaining \emph{a priori} predictions for the BAO observables in each case. After briefly recapping how these should be calculated in a homogeneously expanding cosmology (Section~\ref{BAO_observables_homogeneous}), we explain how a local void distorts the relation between these observables and the redshift (Section~\ref{BAO_observables_void}). We then compile a list of BAO observations from the past twenty years (Table~\ref{BAO_sample_table}). We use this to test the performance of the local void models and the homogeneous \emph{Planck} cosmology adopted in \citetalias{Haslbauer_2020}.

Our results show that $D_{\mathrm{H}}$ and the AP parameter are presently not very discriminative, partly because of observational difficulties, but also because of various effects in the void model cancelling each other out and thereby reducing the expected deviations from the void-free model. Results based on $D_{\mathrm{c}}$ fit somewhat better in the local void models (Figure~\ref{D_c_graph}). We found that the most discriminative quantity is the isotropically averaged BAO distance $D_{\mathrm{V}}$, whose ratio with expectations in the homogeneous model is shown in Figure~\ref{D_V_graph} as a function of $z$. The results generally conform much more closely to the void models than the void-free model, apart from a single data point that is discrepant by close to $3\sigma$ with the former but in just under $1\sigma$ tension with the latter.

We use an approximate $\chi^2$ statistic to estimate the overall goodness of fit of each model to our compilation of BAO observations, removing duplicates where a study reuses data that underpinned an earlier study. The $\chi^2$ values obtained in this way are summarized in Table~\ref{chi_sq_table}, which shows that the void-free model provides the worst fit in nearly all cases. Focusing on $D_{\mathrm{V}}$ due to the larger number of available measurements and their smaller uncertainties, we see that the void-free model has $\chi^2 = 76$, which implies $3.3\sigma$ tension for 42 data points. The void models can reduce this to $47 - 51$ depending on the density profile, even though the parameters in each case were fixed by \citetalias{Haslbauer_2020} without considering BAO datasets. The preference for the void models by $\Delta \chi^2 = 24 - 28$ implies that they are more likely than the void-free model by a factor of $\exp \left( \Delta \chi^2/2 \right) > 10^5$ when considering only the $D_{\mathrm{V}}$ data. Combining these 42 data points with $D_{\mathrm{c}}$ measurements from the 7 studies in Table~\ref{BAO_sample_table} that only report $D_{\mathrm{c}}$ (and thus do not constrain $D_{\mathrm{V}}$) in order to get a more complete picture strengthens the preference for the void models to $\Delta \chi^2 = 36 - 38$, implying a reduction in tension from $3.8\sigma$ to just $1.1\sigma - 1.3\sigma$. The preference is weaker in other common ways of presenting BAO results (Table~\ref{chi_sq_table}) or when using only DESI~DR2 (Section~\ref{DESI_comparison}). We are unable to shed much light on which of the void models performs best. We note that there are arguments against the Maxwell-Boltzmann profile based on the bulk flow curve at $z < 0.083$ \citep{Mazurenko_2024}, despite its somewhat better fit to the BAO data compared to the other profiles.

Our work highlights the importance of BAO observations at $z \la 0.3$ in testing local solutions to the Hubble tension. These observations are challenging due to the large angular size of the BAO ruler, but they are particularly crucial. There is a trade-off between observations becoming easier at high $z$ and the prediction of the local void model that the largest deviations from the homogeneous \emph{Planck} cosmology are to be found at low $z$. Our predictions can give observers an idea of how to weigh the scientific return of a more accurate BAO measurement at higher $z$ compared to a less accurate measurement at lower $z$. They can also provide guidance on which types of BAO measurements to prioritize, since the expected signal is not the same for angular and line of sight measures ($D_{\mathrm{c}}$ and $D_{\mathrm{H}}$, respectively). In the long term, improvements to the accuracy and redshift coverage will allow the AP test, making the results immune to the assumed value of $r_{\mathrm{d}}$ \citep[Section~\ref{AP_test}; see also][]{DESI_2025}.

A local void was proposed long before the Hubble tension based on galaxy number counts in the optical \citep{Maddox_1990, Shanks_1990} and especially in the near-infrared \citep{Keenan_2013}. Outflow from the KBC void can solve the Hubble tension in the semi-analytic models of \citetalias{Haslbauer_2020}, who calibrated their model parameters without considering BAO data. As a result, the predictions of their model for the BAO observables must be considered \emph{a priori} predictions. Our results show that these predictions are fairly successful, with one or other void model achieving a better fit than the homogeneous \emph{Planck} cosmology in every case (Table~\ref{chi_sq_table}). This is especially true when considering $D_{\mathrm{V}}$, for which the quality and quantity of available data are both superior to other common ways of presenting BAO data (Table~\ref{BAO_sample_table} and Figure~\ref{D_V_graph}). We therefore plan to further test the local void scenario by exploring the local velocity field in more detail (Stiskalek et al., in preparation). This will be complemented by a higher redshift test using SNe, which we plan to analyse using a novel method that allows greater flexibility in the relation between luminosity distance and redshift, thereby reducing sensitivity to the assumed cosmological model \citep{Lane_2025, Seifert_2025}.

\section*{Acknowledgements}

IB is supported by Royal Society University Research Fellowship grant 211046 and was supported by Science and Technology Facilities Council grant ST/V000861/1. The contribution of VK was enabled by an undergraduate research bursary from the Royal Astronomical Society. IB thanks Harry Desmond and Pedro Ferreira for helpful discussions. The authors are grateful to the referee for comments which helped to improve this paper.


\section*{Data Availability}

Details of the void simulations were previously published in \citetalias{Haslbauer_2020} along with the parameters of the most likely model for each void density profile. The observational BAO results used here were previously published in the cited studies. An \textsc{excel} summary table with these measurements is available on request to the second author.

\bibliographystyle{mnras}
\bibliography{BAO_bbl}

\begin{appendix}

\section{Impact of a local void on the CMB}
\label{CMB_impact_void}

A major advantage of the \citetalias{Haslbauer_2020} void models is that they preserve the \emph{Planck} background cosmology. While it is necessary to enhance the growth of structure on scales $\ga 100$~Mpc to explain the observed KBC void, the physical scales in the early Universe were much smaller, so it is quite possible that whatever new physics is required primarily affects the late universe. Without any modification to $\Lambda$CDM physics prior to recombination or to the cosmic expansion history, the \citetalias{Haslbauer_2020} models would have little impact upon the CMB anisotropies.

There are a few subtler ways in which the void scenario would nonetheless impact the CMB. Our peculiar velocity in the CMB frame would contribute to the CMB dipole, placing constraints on our vantage point within the void \citepalias[see figure~8 of][]{Haslbauer_2020}. If we are not exactly at its centre, there would also be a contribution to the CMB quadrupole from gravitational lensing by the local void \citep*{Alnes_2006, Nistane_2019}. However, this effect would be very small for realistic vantage points (Stiskalek et al., in preparation). Higher-order multipoles would be even less affected. Similarly, the proposed $\approx 20\%$ underdensity out to 300~Mpc is well within current kSZ bounds \citep{Ding_2020, Cai_2025}.

A local void would create a local hill in the gravitational potential $\Phi$. The resulting distortion to spacetime would slightly impact $D_{\mathrm{c}}^{\mathrm{CMB}}$, the comoving distance to the CMB. To estimate this, we note that GR would globally redshift the CMB by 0.8\% \citepalias[section~5.3.3 of][]{Haslbauer_2020}, which those authors argued lies well within current uncertainties \citep{Yoo_2019}. This implies the general relativistic line element is distorted by 0.8\% at the void centre, which we can approximately take to be our vantage point. The distortions would have been much smaller along the majority of our past lightcone because $\Phi$ would be smaller further from the void centre and further back in time, when the void was shallower. Assuming that $\Phi$ decays along our past lightcone with an exponential scale length of 700~cMpc \citep[far larger than the 300~Mpc scale of the KBC void;][]{Keenan_2013}, the void-induced distortion to spacetime would effectively apply to only 5\% of the 14~cGpc journey of a CMB photon. This implies the local distortion to spacetime due to the KBC void's gravity would affect $D_{\mathrm{c}}^{\mathrm{CMB}}$ by $\la 0.04\%$, which is far smaller than the 0.2\% uncertainty on the \emph{Planck} value of $r_{\mathrm{d}}$. It is therefore clear that the small expected change to $D_{\mathrm{c}}^{\mathrm{CMB}}$ due to spacetime distortion can easily be compensated by very slight adjustments to the cosmological parameters. In fact, we do not even need to adjust the parameters at all -- changing $D_{\mathrm{c}}^{\mathrm{CMB}}$ by only 0.04\% would affect the angular scale of the first peak in the CMB power spectrum by the same amount, which only slightly exceeds the observational uncertainty of 0.025\% \citep[see table~5 of][]{Tristram_2024}. Therefore, it is reasonable to assume that the \citetalias{Haslbauer_2020} void models only negligibly affect the excellent $\Lambda$CDM fit to the CMB temperature and polarization power spectrum across a very wide range of angular scales \citep{Calabrese_2025}.

\end{appendix}

\bsp 
\label{lastpage}
\end{document}